\documentclass[a4paper,usenatbib]{mnras}

\usepackage[T1]{fontenc}
\usepackage{ae,aecompl}

\usepackage{graphicx}	
\usepackage{amsmath}	
\usepackage{amssymb}	

\title[HI-selected galaxies in hierarchical models]{
HI-selected Galaxies in Hierarchical Models of Galaxy Formation and Evolution.}

\author[A. Zoldan et al.]{
Anna Zoldan,$^{1,2}$\thanks{E-mail: zoldan@oats.inaf.it}
Gabriella De Lucia,$^{2}$
Lizhi Xie,$^{2}$
Fabio Fontanot$^{2}$ and
\newauthor 
Michaela Hirschmann$^{3}$
\\
$^{1}$Physics Department, Universit\'a degli Studi di Trieste, Via Valerio 2, 34127-Trieste, TS, Italy\\
$^{2}$OATS, INAF, Via Bazzoni 2, 34124-Trieste, TS, Italy\\
$^{3}$ Institut d'Astrophysique de Paris, Sorbonne Universit\'{e}s, UPMC-CNRS, UMR7095, F-75014, Paris, France
}

\date{Accepted XXX. Received YYY; in original form ZZZ}

\pubyear{2015}

\begin{document}
\label{firstpage}
\pagerange{\pageref{firstpage}--\pageref{lastpage}}
\maketitle

\begin{abstract}
In this work, we study the basic statistical properties of HI-selected galaxies
extracted from six different semi-analytic models, all run on the same
cosmological N-body simulation. 
One model includes an explicit treatment for the partition of cold 
gas into atomic and molecular hydrogen. 
All models considered agree nicely with the measured HI mass function
in the local Universe, with the measured scaling relations between HI and
galaxy stellar mass, and with the predicted 2-point correlation function for 
HI rich galaxies. 
One exception is given by one model that predicts very little HI associated 
with galaxies in haloes 
above ${\sim}10^{12}{M}_{\odot}$: we argue this is due to a too
efficient radio-mode feedback for central galaxies, and to a combination of
efficient stellar feedback and instantaneous stripping of hot gas for satellites. 
We demonstrate that treatment of satellite galaxies introduces large
uncertainties at low HI masses. 
While models assuming non instantaneous stripping of hot gas tend to form 
satellite galaxies with HI masses slightly smaller than those of centrals with 
the same stellar mass, instantaneous gas stripping does not translate necessarily 
in lower HI masses. 
In fact, the adopted stellar feedback and star formation affect the satellites too. 
We analyze the relation between HI content and
spin of simulated haloes: low spin haloes tend to host HI poor galaxies while
high spin haloes are populated by galaxies in a wide range of HI mass. In our
simulations, this is due to a correlation between the initial gas disk size and
the halo spin. 

\end{abstract}



\section{INTRODUCTION}
Cold gas has a central role in galaxy evolution as it is involved in
virtually all processes at play: cooling, star formation,
feedback, mergers and processes related to the `environment' such as
ram-pressure and strangulation.
The main component of cold gas is neutral hydrogen, in its atomic and
molecular forms, each distributed in a `disk' with a specific density
profile \citep{leroy2008}.  Observationally, HI can be observed
directly through the 21~cm line, while ${\rm H}_2$ must be inferred
from the $^{12}$CO content through a factor, $\alpha _{CO}$, whose
value and dependence on other physical properties (in particular the
gas metallicity) are not well known
\citep{magdis2011,bolatto2011,leroy2011,narayanan2012,hunt2015,amorin2016}.

Surveys such as the HI Parkes All Sky Survey
\citep[HIPASS,][]{zwaan2005} or the Arecibo Legacy Fast ALFA survey
\citep[ALFALFA,][]{martin2010}, have recently collected large amounts of data,
but since the HI signal is relatively faint, these data are generally
limited to low redshift and relatively HI rich galaxies.  These
observations have provided detailed information on the basic statistical
properties of HI selected galaxies in the local Universe, in
particular the HI mass function \citep{zwaan2005,martin2010} and the
2-point correlation function
\citep{meyer2007,martin2012,papastergis2013}.  We live in a very
exciting era for HI observations, as new radio facilities such as the
Square Kilometre Array (SKA) and its precursors/pathfinders will soon
extend significantly the redshift range and dynamical range in HI mass probed
\citep{obreschkow2009,kim2011}, providing a new crucial testbed for
our models of galaxy formation and evolution. It is therefore
important to develop our theoretical tools, and assess how their basic
predictions compare with the available data. 

Until recently, the inter-stellar medium (ISM) of galaxies, in both
semi-analytic models of galaxy formation and hydrodynamical simulations, was
treated as a single phase constituted of gas with temperature below
$10^4$~K. Stars were assumed to form from this gaseous phase adopting different
parametrizations, generally based on some variation of the Kennicutt-Schmidt
law \citep{kennicutt1989}. In the last decade, our understanding of how star
formation depends on the local conditions of the ISM has improved significantly
thanks to new high-resolution observations of the multiphase gas in samples of
nearby galaxies. These data have shown that the density of star formation
correlates strongly with the molecular gas density, while there is almost no
correlation with the total gas density \citep[e.g.][and references
  therein]{blitz2006,leroy2008}. These results have triggered significant
activity in the theoretical community aimed at including an explicit treatment
for the transition from atomic to molecular hydrogen, and molecular hydrogen
based star formation laws \citep{fu2010mol,lagos2011mol, christensen2012mol,
  kuhlen2013mol}. Recent studies have
focused on the impact of different physical processes and of cosmology on the
HI distribution and properties of HI selected galaxies
\citep[e.g.][]{popping2009,obreschkow2009,kim2011,dave2013,rafie2015,kim_hs2016,crain2016}.

In this paper, we focus on results from semi-analytic models of galaxy
formation. While these methods do not include an explicit treatment of
the gas dynamics and prevent (at least in their standard
implementations) studies of spatially resolved properties of galaxies,
their computational cost is typically significantly lower than that of
high-resolution hydrodynamical simulations. They therefore offer a
flexible and efficient tool to analyze the dependence of model
predictions on specific assumptions and parameterizations. We focus
here on the basic statistical properties of HI selected galaxies
(i.e. their distributions in mass, their halo occupation distribution,
and their 2-point clustering function). To analyze the dependence on
the different modeling adopted, we use four different models
whose results are publicly available on the Virgo-Millennium Database
\citep{lemson2006}, and other two models recently developed by our
group \citep[][and Xie et al. in preparation]{hirschmann2015}. The
latter model, in particular, includes an explicit treatment for
partitioning the cold gas associated with model galaxies in HI and ${\rm
  H}_2$. For 
all other models, as we will discuss below, the HI content of model
galaxies has been derived post-processing model outputs.

The layout of the paper is as follows. In Sec.~\ref{sec:SAM}, we
briefly describe the models used in our study, their backbone
simulations, and the algorithm we use to build mock light-cones from
model outputs.  In Sec.~\ref{sec:HI_extra_rel}, we explain how we
assign HI masses to model galaxies where this is not an explicit model
output, and compare the distribution of HI masses and basic scaling
relations predicted by the models with available observational data.
In Sec.~\ref{sec:2PCF_theory_models}, we analyze the predicted 2-point
correlation function of HI-selected galaxies, and compare model
predictions with recent measurements by \citet{papastergis2013}. We
discuss the results on the basis of the predicted halo occupation
distribution.  In Sec.~\ref{sec:results_HI_satellites}, we discuss the
role of satellite galaxies and study the predicted evolution of their
HI content. In Sec.~\ref{sec:HI_DM}, we study the relation between the
HI in galaxies and the physical properties of the hosting dark
matter haloes. Finally, in Sec.~\ref{sec:conclusions}, we summarize our
results and give our conclusions.

\section{SIMULATIONS AND GALAXY FORMATION MODELS}
\label{sec:SIMSAM}

In this work, we take advantage of 6 different semi-analytic models of galaxy
formation and evolution, that originate from the work of three independent
research groups.  All models are run on the same N-body cosmological
simulation, i.e. the Millennium Simulation (Sec.~\ref{sec:MS}), and all follow the principles
outlined in \citet{wf91} including specific (different) modeling for cooling,
star formation, stellar feedback, mergers and starbursts, disk instabilities,
chemical enrichment and feedback from Active Galactic Nuclei (AGN). 
Using models that employ different prescriptions for processes
involving cold gas, it is possible to quantify the relative importance 
in determining the observed relations. At the same time, systematic 
(i.e. common to all models) disagreements can be used to identify 
specific aspects of the models that need to be improved.
Below, we
briefly outline the main differences between the models considered, the main
characteristics of the simulations used, and describe the software built to
construct mock light-cones from the available galaxy catalogues.

\subsection{Galaxy formation models}
\label{sec:SAM}

The \citet[][B06 hereafter]{bower06} model has been
developed by the `Durham group' and is an extension of the {\sc GALFORM}
model published in \citet{cole2000} and \cite{benson03}. The model published in
\citet[][hereafter DLB07]{delucia07} has been developed by the `Munich group'
and represents an extension of the model described earlier in
\citet{springel2001}, \citet*{delucia04}, and \citet{croton2006}.  These two
models have been the first whose results were made publicly available through a
relational database\footnote{http://wwwmpa.mpa-garching.mpg.de/millennium/}
that we have heavily used for our study. 

It was early realized that both these models tend to over-predict the number
densities of galaxies smaller than the Milky Way, and the overall fraction of passive galaxies
- problems that have turned to be of difficult solution and shared by all
hierarchical models published in the last years, including hydrodynamical
cosmological simulations \citep*[see e.g.][and references
  therein]{delucia2014r}. The other models we use in this study represent two
independent branches, both based on subsequent upgrades of the DLB07 model, 
that provide significant improvements on these problems. 

The model described in \citet[][G11]{guo10} differs from the DLB07 model 
primarily for a more
efficient feedback and a non-instantaneous stripping of the hot reservoir
associated with galaxies at the time of infall on larger systems. The modified
treatment of satellite galaxies improves the agreement with observational
data as for the fraction of active galaxies, while the more efficient feedback
brings the predicted mass function in agreement with data in the local
Universe. The model, however, still suffers of an excess of intermediate to
low-mass galaxies at higher redshift. 
\citet{henriques2013} 
adopt the same physical model as in G11, but include a variation with cosmic
time and halo mass of the efficiency with which galactic wind ejecta are 
re-accreted. They then use Monte Carlo Markov Chain
methods to identify the parameter space that allows the measured evolution of 
the galaxy mass (and luminosity) function from z=0 to z=3 to be
reproduced. \citet[][H15]{henri15} have later extended this work to the Planck
first-year cosmology, 
by ``re-scaling'' the Millennium Simulation merger trees, as explained  
in Sec.~\ref{sec:MS}. In this work, we use model outputs based on the re-scaled
simulation.

The other independent branch we use in our study has been developed by the
`Trieste group', and is provided by the recently published GAlaxy Evolution and
Assembly ({\sc GAEA}) model \citep{hirschmann2015}, and by an extension of this
model (Xie et al. in preparation, XBR16). The {\sc GAEA} model differs from
DLB07 primarily for the inclusion of a sophisticated chemical enrichment scheme
that accounts for the finite lifetimes of stars and independent yields from
massive stars and both SNII and SNIa, and for an updated stellar feedback
scheme, based on the results obtained in the framework of the Feedback In
Realistic Environments (FIRE) simulations \citep{Hopkins_etal_2014}. The GAEA
model includes modifications of the re-incorporation rate of ejected gas
similar to those suggested by \citet{henri15}. We refer to the original paper
for more details. It is worth stressing that {\sc GAEA} adopts an instantaneous
stripping of the hot gas reservoir associated with infalling galaxies. Yet,
\citet{hirschmann2015} demonstrate that this model is successful in reproducing
simultaneously the evolution of the galaxy stellar mass function, the fraction
of passive galaxies\footnote{The quiescent fraction measured for the lowest mass galaxies
  is still larger than observational estimates, but improved significantly with
  respect to predictions from the DLB07 model.}  as a function of stellar mass
observed in the local Universe, and the observed evolution of the relation
between galaxy stellar mass and metallicity content of the gaseous phase.

All models introduced so far treat the gas as a single phase component, and
therefore need to be {\it post-processed} to infer the HI content associated
with each model galaxy, as described in Sec.~\ref{sec:HI_pres}. The XBR16 model
represents an update of the GAEA model including a treatment for the partition
of the gaseous phase in atomic and molecular gas, based on the empirical
relation by \citet{blitz2006}. Therefore, among all models considered in our
study, this is the only one including a self-consistent treatment of the star
formation rate, which depends on the molecular hydrogen content of galaxies.
In this model the disk is divided in 21 concentric annuli (in all other models
star formation is modelled assuming a single value for the entire disk), and
the star formation rate is estimated in each of the annuli assuming it is
proportional to the surface density of molecular gas. A partition of the cold gas into
its molecular and atomic components is performed at each time-step (just before
star formation takes place) of the simulation, in each annulus. This allows
star formation to continue for longer times than in the other models in
the inner parts of the disk. This more sophisticated treatment of the star
formation process leads to relevant differences with respect to the GAEA model, as we
will show in the next sections.

Among the models used in this paper, B06, DLB07, G11 and H15 are 
publicly available at the relational database mentioned above, developed 
as part of the activities of the German Astrophysical Virtual Observatory (GAVO).
The GAEA and XBR16 models are not yet public, but will soon be released 
on the same database. Additional models, including more recent versions of
the {\sc GALFORM} model featuring a treatment for the partition of cold gas in atomic 
and molecular hydrogen, are accessible form an alternative database server 
at the Institute for Computational 
Cosmology\footnote{http://galaxy-catalogue.dur.ac.uk:8080/Millennium/}.

\subsection{Dark matter simulation and merger trees}
\label{sec:MS}

All models presented in the previous section have been run on the Millennium
Simulation \citep{springel2005}. This is a cosmological N-body dark matter only
simulation, that follows the evolution of $N=2160^3$ particles of mass
$8.6\times10^{8}\,h^{-1}{\rm M}_{\odot}$ within a comoving box of
$500\,h^{-1}{\rm Mpc}$ on a side, and with a cosmology consistent with WMAP1
data. In particular, the values of the adopted cosmological parameters are:
$\Omega_{b}=0.045$, $\Omega_{m}=0.25$, $\Omega _{\Lambda}=0.75$,
$H_0=100h\,{\rm Mpc}^{-1}\,{\rm km}\,{\rm s}^{-1}$, $h=0.73$, $\sigma_{8}=0.9$
and $n_s=1$.  Simulation data were stored at 64 output redshifts, each
corresponding to a snapshot number $n$ (with $n$ varying from 0 to 63)
through the following formula: $\log(1+z_{n})=n(n+35)/4200$. 

Dark matter haloes were identified using a standard friends-of-friends algorithm
with a linking length of $0.2$ in units of the mean particle separation. The
algorithm SUBFIND \citep{springel2001} was then used to identify bound
substructures with a minimum of 20 particles, which corresponds to a halo mass
resolution limit of $M_h=1.7\times10^{10}{\rm M}_{\odot}h^{-1}$.  Halo merger
trees were built tracing the majority of the most bound particles of each
subhalo from a given snapshot to the following one. In this way, each halo was
assigned a unique descendant. 

As mentioned above, H15 is based on a different cosmology: the authors still
use the Millennium Simulation but `rescale' it to the PLANCK cosmology
\citep{planck2014} using the technique discussed in \citet{angulo2010}, as
updated by \citet{angulo2015}. Specifically, the adopted cosmological
parameters are: $\Omega _{b}=0.0487\;(f_b=0.155)$, $\Omega _{m}=0.315$, $\Omega
_{\Lambda}=0.685$, $h=0.673$, $\sigma_{8}=0.829$ and $n_s=0.96$.  We note that
as found in previous studies \citep[e.g.][]{wang_jie_2008,guo_2013_wmap},
relatively small variations of the cosmological parameters are found to have
little influence on the overall model predictions once model parameters are
(slightly) modified to recover the same model normalization.

To control the resolution limits of our models, we have also taken advantage 
of the Millennium II simulation
\citep[][MS-II]{boylan-kolchin2009}. The size of this simulation is one fifth
that of the Millennium Simulation (i.e. 100 Mpc $h^{-1}$ on a side), but the
particle mass is 125 times lower, resulting in a minimum resolved halo mass  of 
$M_h=1.4\times10^{8}{\rm M}_{\odot}h^{-1}$. 
Below, we will use G11, H15 and XBR16 outputs based on the MS-II in order to check the 
reliability
of some relations, in particular those involving dark matter halo masses in the 
$10^{11}-10^{12}\;M_{\sun}h^{-1}$ range.

\subsection{Light-cones algorithm}
\label{sec:lightcone}

In Sec.~\ref{sec:2PCF_theory_models} we analyze the 2-Point Correlation Function
of galaxies selected according to their HI content. We use mock light-cones and
the projected correlation in order to carry out a straightforward  comparison with observations.
In addition, the construction of many light-cones (some dozens) allows us to give
an estimate of the error on the projected 2PCF due to the cosmic variance, as well as of
the random noise expected for the real 2PCF. 

We have developed an algorithm that creates mock catalogues from the outputs of
galaxy formation models.  We use these catalogues in the calculation of the
2PCF (Sec.~\ref{sec:2PCF_theory_models}).  The approach is very similar to that
adopted in the Mock Map Facility developed by \citet[][MoMaF]{blaizot2005}, and
we briefly summarize it below. We refer to the original paper for more details.

As explained above, the outputs of the galaxy formation models used in this
study are given by galaxy catalogues stored at a finite number of
snapshots, each corresponding to a different redshift $z_i$. Galaxy catalogues
contain a number of physical properties (i.e. masses, metallicities,
luminosities, etc.) and consistent redshift and spatial information, i.e. the
position of each model galaxy within the simulated box, as well its
velocity components. The first problem that arises when constructing mock
observations from these types of outputs is that redshift varies continuously
along the past light cone while outputs are stored at a finite number of
redshifts (in our particular case spaced at approximately 300 Myr intervals out
to $z=1$). The other problem arises because of the need to `replicate' the same
simulated region of the Universe (in our case corresponding to a box of
$500\,{h}^{-1}\,{\rm Mpc}$ comoving on a side) several times to fill the
light-cone.

\begin{figure}
  \includegraphics[trim=1cm 1cm 1cm 1cm, clip, width =
    0.9\columnwidth]{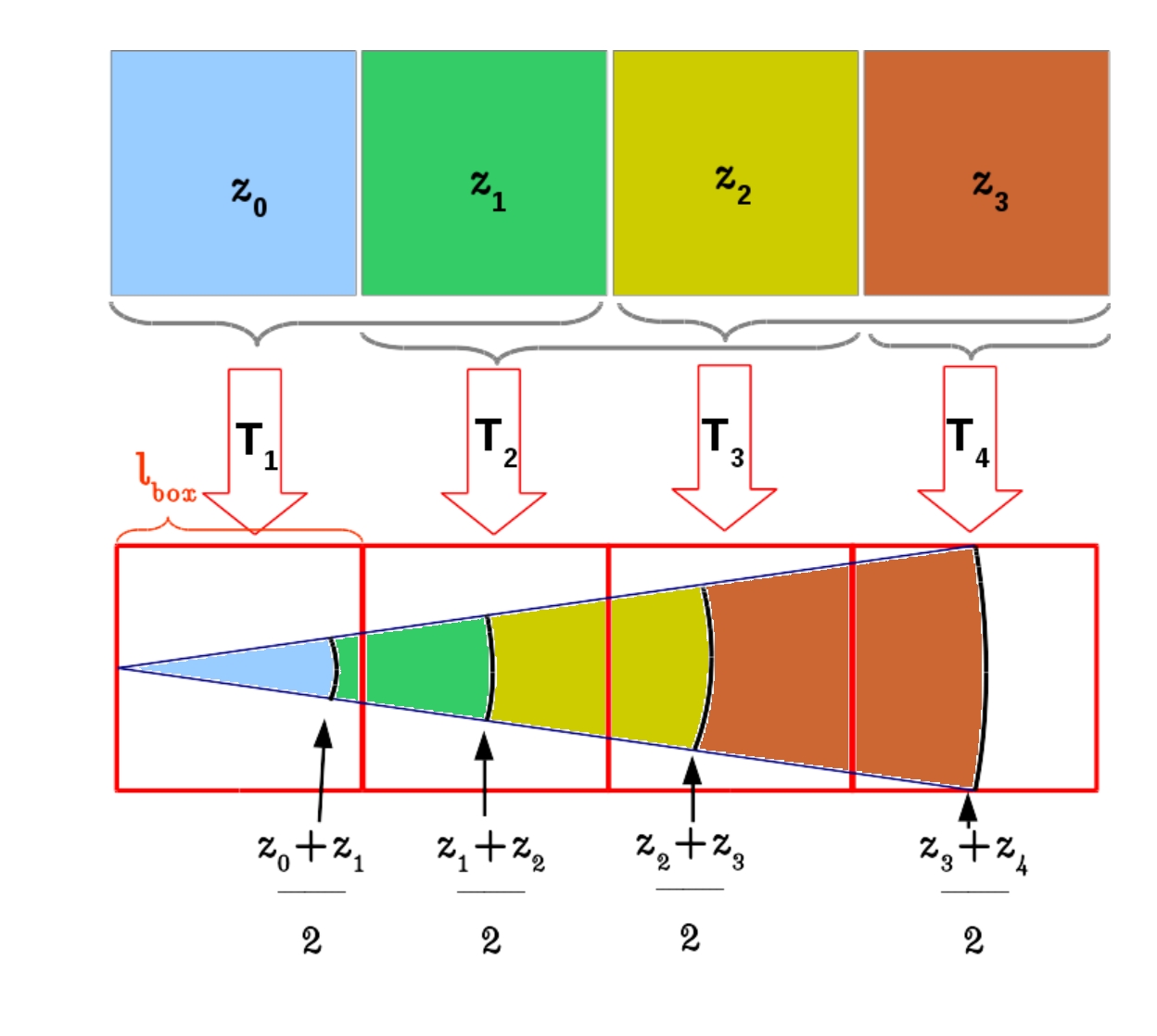} 
  \caption{A schematic illustration of our light-cone construction
    algorithm: a grid of cubic boxes with size length $l_{box}$ (red)
    is built over the cone, and galaxies inside each box of the grid
    are randomly shifted/rotated/inverted depending on the specific
    box they end up in.  Galaxies are extracted from the simulation
    output with redshift closest to the co-moving distance to the
    observer.}
  \label{fig:lightcone}
\end{figure}

We try to minimize these problems by building the light-cone as follows (we 
refer to Fig.~\ref{fig:lightcone} for a schematic description of our 
method).  
\begin{enumerate}
 \item We build a 3-D grid made of cubes of the same size $l_{box}$ of
   the simulation box (the red grid in the figure).
 \item The cone is placed inside this grid horizontally and moved so
   as to overlap with the smallest number of grid boxes (the blue cone
   in the figure).
 \item The cone is divided into different regions whose edges are
   given by the average redshift of two subsequent snapshots, namely
   $z^{lc}_{i}=(z_i+z_{i+1})/2$ (black arrows in the figure).
 \item Model galaxies are then placed inside the cone according to the
   following rules:
	\begin{itemize}
	 \item Each redshift region $z^{lc}_i$ of the cone is filled
           with galaxies from the box corresponding to the closest
           redshift $z_i$ (the correspondence is colour-coded in the
           figure).
	 \item The positions (and velocities) of model galaxies are
           randomly transformed, in order to avoid replications of
           structures along the line of sight. A specific random
           transformation is assigned to each 3-D grid box.  This is a
           combination of a random shift, a random rotation of 0,
           $\pi/2,\,\pi$ or $3\pi/2$ around a random axis or an
           inversion of coordinates along a random axis.  In the
           figure, a specific transformation is indicated with a big
           arrow.
	\end{itemize}

\end{enumerate}

Referring to the specific example shown in Fig.~\ref{fig:lightcone}, the 
lightcone element corresponding to the second grid box (from the left) is 
divided in two redshift regions by the edge $z^{lc}_2 = (z_1 + z_2)/2$. The 
positions (and velocities) of the galaxies in the snapshots corresponding to 
$z_1$ and $z_2$ are transformed using the same specific transformation $T_2$. 
The transformed snapshots are then used to fill the grid element considered. 
Galaxies below  $z^{lc}_2$ are extracted from the transformed snapshot $z_1$, 
while those above $z^{lc}_2$ are taken from the transformed snapshot $z_2$.

Adopting this approach, large scale structures are affected at the edges of the
3D grid boxes, but not at the redshift separation between subsequent
snapshots. This allows accurate measurements of the 2-point correlation
function (2PCF). 

\citet{blaizot2005} carried out a detailed analysis of replication effects on
the clustering signal. In particular, they quantified the effect due to (i)
random tiling and (ii) finite volume. The random tiling approach introduces a
negative bias in the mock catalogues because it decorrelates pairs of galaxies
when reshuffling them to suppress replication effects. \citet{blaizot2005} gave
a theoretical estimate of the relative error in the number of pairs due to
shifting of boxes. For the spatial correlation function, this is found to be
less than 10 per cent on scales ranging from 1 to $\sim 10\,{h}^{-1}\,{\rm
  Mpc}$. A numerical estimate was also computed by subdividing the volume of
the simulation used in their study (a cube of $100\,h^{-1}_{100}\;Mpc$ on a side) in
$8^3$ sub-volumes to which translations, rotations and inversions were applied
as described above. We have repeated the same exercise by taking advantage of
the larger volume of the Millennium Simulation and considering sub-boxes of
different sizes ($L_{box}=250,\; 125\; Mpc/h$). For the analysis presented in
the following, light-cones are built starting from the entire volume of the 
MS ($L_{box}=500\; Mpc/h$), and we have verified that the random tiling 
approach is not expected to significantly affect our measurements. 

The finite volume of the simulation prevents us from studying fluctuations 
larger than $\sim V^{1/3}$. Since in this study we focus on clustering on 
relatively small scales ($\sim 20\,h^{-1}\,{\rm Mpc}$), we do not expect the 
finite volume of our simulation to affect significantly our measurements.  

\section{Neutral hydrogen distribution and scaling relations}
\label{sec:HI_extra_rel}
In this section, we study the distribution of HI masses and basic scaling
relations predicted by our models, and compare them with recent observational
determinations. Given the resolution mass limit of the Millennium Simulation,
all the following analysis is limited to galaxies with 
$M_*>10^9\,{\rm M}_{\odot}$.
 Below this limit the sample is incomplete (this is evaluated comparing the 
galaxy stellar mass function for models based on the MS and on the MS-II).
Using the HI mass function, Xie et al. have determined the completeness 
limit in HI mass, that is assessed to be $M_{HI} \sim 10^{8.8}M_{\odot}$. 
In the following analysis, we study relations involving HI masses 
slightly below this limit. 
We checked the influence of incompleteness using the MS-II for the G11, H15 and 
XBR16 models, and found that, qualitatively, our results do not change.

The following analysis is performed on a sub-volume of the MS of one sixth, 
but we have verified that results do not change significantly when considering 
a larger volume.

\subsection{Neutral hydrogen mass estimates}
\label{sec:HI_pres}

XBR16 is the only model among those considered in our study that accounts
self-consistently for a direct dependence of star formation rate on the
molecular hydrogen density \citep{blitz2006,leroy2008}. This is done by
introducing a partition of the cold gas phase in atomic and molecular hydrogen,
based on the empirical relation by \citet{blitz2006}, at the time of each star
formation episode.  Therefore, the HI and ${\rm H}_2$ content of model galaxies
are direct outputs of this model, even if they are not explicitly followed as
two separated components.  For all the other models, star formation is assumed
to depend on the total cold gas content through a classical
\citet{kennicutt1989} relation. In this case, the HI content of galaxies can be
estimated using two different methods:
\begin{itemize}
  \item Combining the observed gas density profile \citep{leroy2008} and the 
    dependence of $R_{mol}=M_{H_2}/M_{HI}$ on the disk mid-plane pressure
     \citep{blitz2006}, \citet{obreschkow2009} find the following
    relation: 
    $$R_{mol}=[3.44R_c^{-0.506}+4.82R_c^{-1.054}]^{-1}.$$ In this equation,
    $R_c\sim[r_{disk}^{-4}M_{cold}(M_{cold}+0.4M_*)]^{0.8}$ is the central 
    $R_{mol}$, $r_{disk}$ is the disk scale length, $M_{cold}$ the cold gas
    mass in the disk, and $M_*$ the stellar mass of the disk.
  \item Following \citet{baugh2004} and \citet{power2010}, we can assume a
    direct proportionality between the cold gas content and HI.
    Helium amounts to about 24 per cent of the total cold gas, while metals
    represent a negligible fraction. Assuming the HII content is negligible,
    hydrogen can be divided in $\sim 71$ per cent of HI and $\sim 29$ per cent
    of H$_2$, namely $R_{mol}=M_{H_2}/M_{HI}=0.4$
    \citep{keres2003,zwaan2005}. 
\end{itemize}

\citet{power2010} analyzed the impact of these two approaches on different
semi-analytic models (including some used in our study) and found little
differences between them at low redshift (differences are larger at
high redshift). We confirm that there are only small differences between the HI
content of model galaxies estimated using the two methods described above, for 
all models used in our study. Therefore, unless otherwise stated, we will 
always compute HI using the \citet{obreschkow2009} prescription.

\subsection{Stellar and HI Mass Functions}
\label{sec:MF}

\begin{figure}
  \centering
  \includegraphics[trim=1cm 0.4cm 1cm 1.5cm, clip, width =
    \columnwidth]{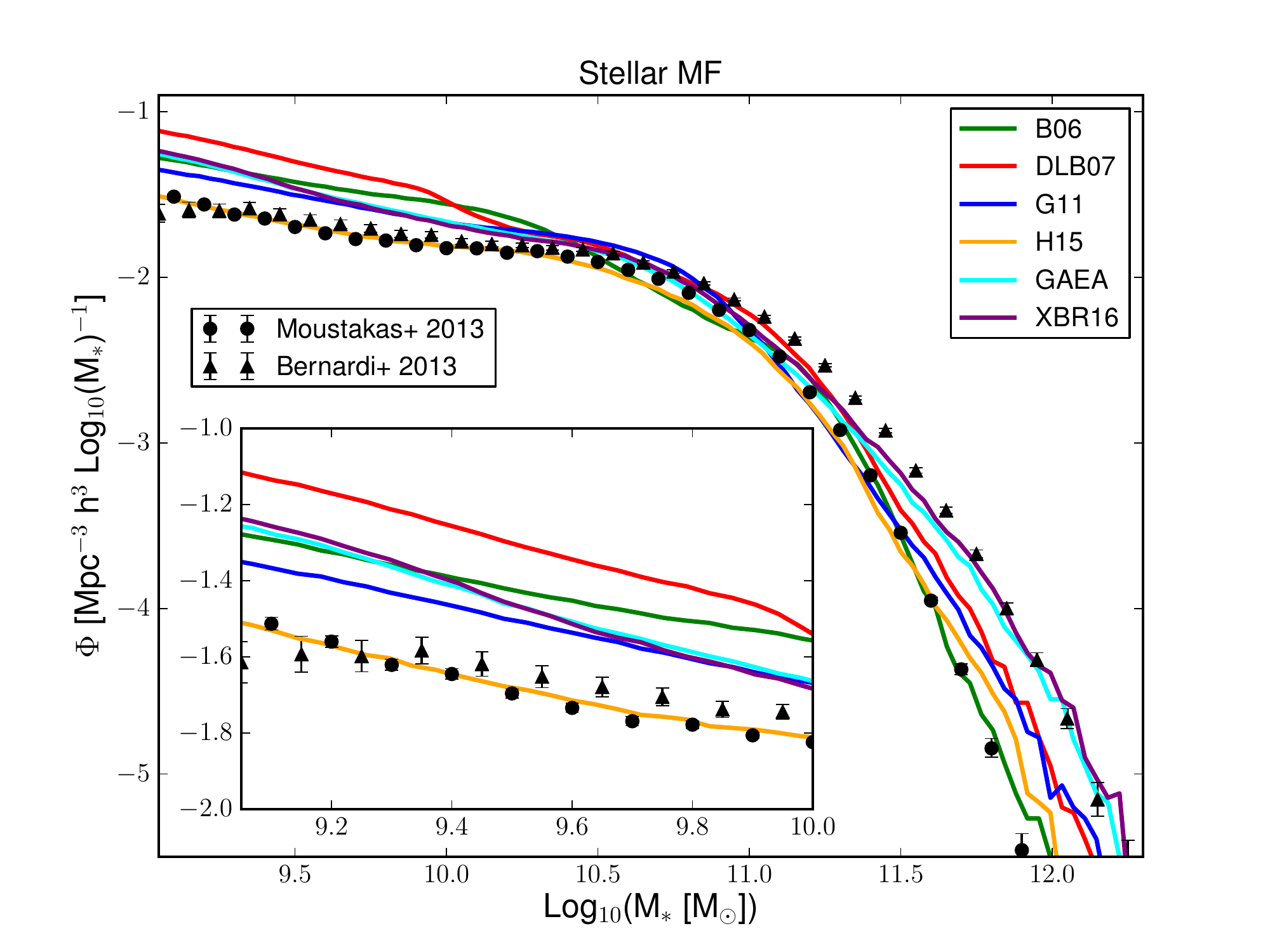} 
  \caption{\label{fig:MF_star} The galaxy stellar mass function predicted by
    the six models used in our study, compared to observational measurements by
    \citet[][circles]{moustakas2013} and \citet[][triangles]{bernardi2013},
    both based on SDSS. The inset shows a zoom on the low-mass end.
   }
 \end{figure}

In Fig.~\ref{fig:MF_star}, we compare the galaxy stellar mass function
predicted by the models with observational measurements by
\citet{moustakas2013} and \citet{bernardi2013}. Both determinations are based
on SDSS data, but assume different estimates of stellar mass. In particular,
the latter is based on a S\'ersic-exponential fit to the surface brightness
profiles of galaxies, and translates in significantly higher number densities
of massive galaxies. The two measurements highlight that, while uncertainties
are relatively small for the intermediate and low-mass regime, they are much
larger above the knee of the mass function. Furthermore, in this mass regime
also statistical uncertainties are larger because of the small number of
galaxies.

All the models reproduce well the observational measurements around the knee,
with a slight underestimation for the B06 and H15 models. The low-mass end of
the mass function is well reproduced only by the H15 model, while the largest
over-prediction is found for the DLB07 and B06 models. As mentioned above, the
good agreement found for the H15 model is obtained by construction as these
authors tune their model parameters to reproduce the evolution of the galaxy
mass function from z=0 to z=3. At high masses, models tend to be closer to the
\citet{moustakas2013} determination up to stellar masses $\sim 10^{11}\,{\rm
  M}_{\odot}$, while they basically cover all the range between the two
measurements shown for larger masses. For the most massive galaxies, the lowest
number densities are obtained by the B06 and H15 models, the largest by GAEA 
and XBR16. 

\begin{figure}
  \centering
  \includegraphics[trim=1cm 0.4cm 1cm 1.5cm, clip, width =
    \columnwidth]{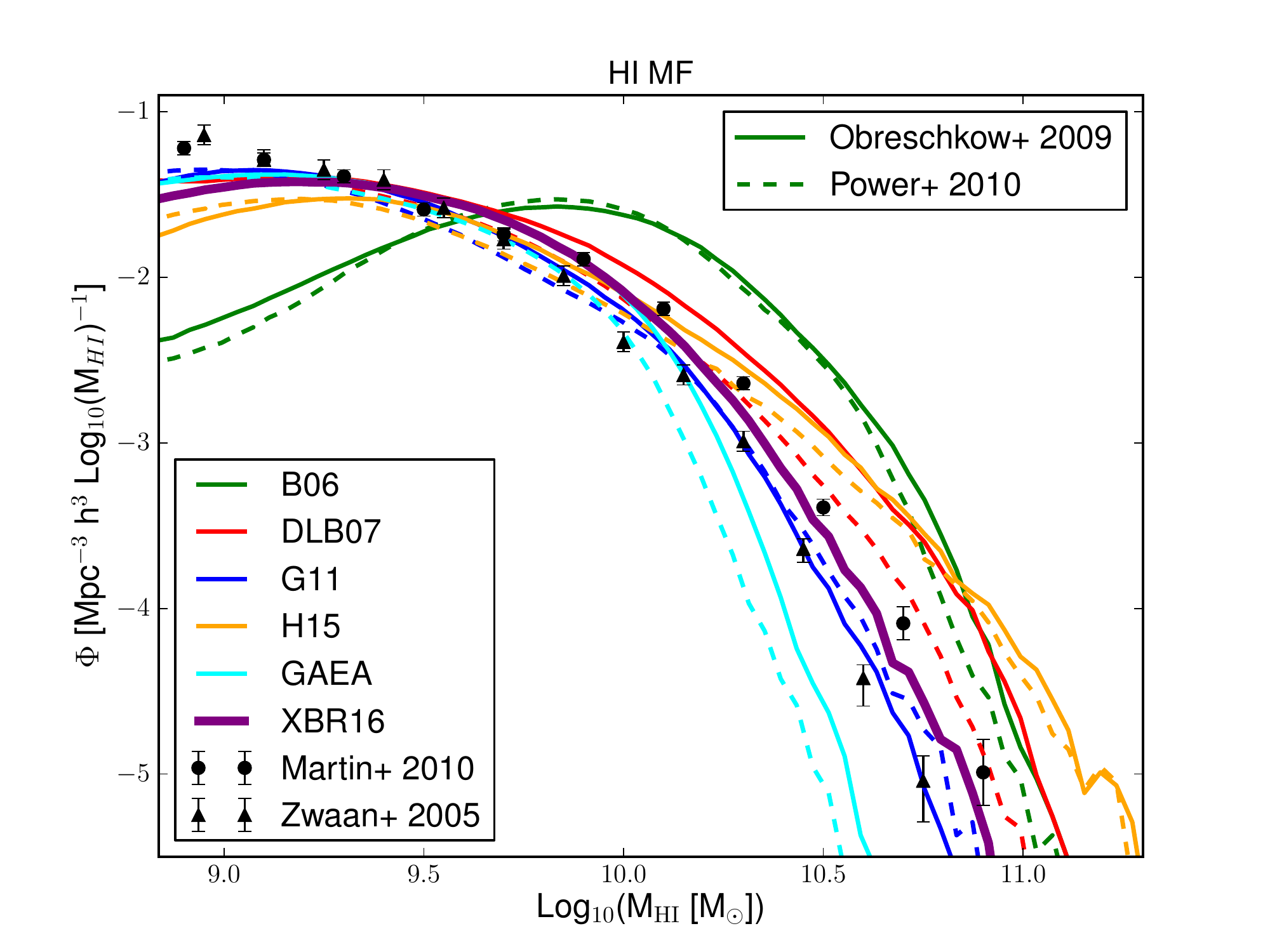} 
  \caption{\label{fig:MF_HI} The HI mass function predicted by the models,
    compared to observational data by \citet[][circles]{martin2010} and
    \citet[][triangles]{zwaan2005}. Solid lines correspond to the
    \citet{obreschkow2009} method to assign HI masses, while dashed lines
    correspond to the assumption of a constant proportionality between total
    cold gas and HI mass as in \citet{power2010}. 
     }
 \end{figure}

Fig.~\ref{fig:MF_HI} shows the HI mass function predicted by the models used in
our study, obtained using the methods discussed in Sec.~\ref{sec:HI_pres} (with
the exception of XBR16, that returns a direct estimate of the HI mass). Model
predictions are compared to observational data from \citet{zwaan2005}, based on 
HIPASS, and from \citet{martin2010}, based
on the ALFALFA survey. 

The two different HI estimates (solid and dashed lines in the figure) return
quite similar results, with the \citet{power2010} prescription slightly
under-predicting the HI high mass end with respect to the method suggested by
\citet{obreschkow2009}. 
 
B06 is the model that deviates most from the data. In particular, it
over-predicts the number densities of galaxies with high HI masses
($M_{HI}\geq10^9M_{\odot}h^{-2}$), and under-predicts it for lower masses.
\citet{kim2011} already discussed this limitation of B06 and proposed a new
version of this model tuned to reproduce the HIMF of \citet{zwaan2005}. This
modified model also resulted in slightly better agreement with the observed
2PCF based on HIPASS. G11 and XBR16 reproduce quite well the observations,
while the GAEA model underestimates the high mass end. We remind that GAEA and
XBR16 are based on the same physical model but the latter includes an explicit
treatment for the partition of cold gas in atomic and molecular hydrogen in
disk annuli, and a star formation law that depends on ${\rm H}_2$. The division
in annuli translates in the XBR16 model having less H$_2$ at fixed star
formation rate, and this leads to a larger amount of HI (at fixed cold
gas). This effect is enhanced for gas rich galaxies.  Finally, we note that
the XBR16 model is tuned to reproduce the HIMF.
 
The slight underestimation of the HIMF at the low mass end in all the models is due to
resolution (Xie et al., in preparation).
   
\subsection{Scaling relations}
\label{sec:HI_Ms}

\begin{figure*}
    \includegraphics[trim=0.5cm 0.8cm 0.5cm 0.8cm, clip,
      width=\textwidth]{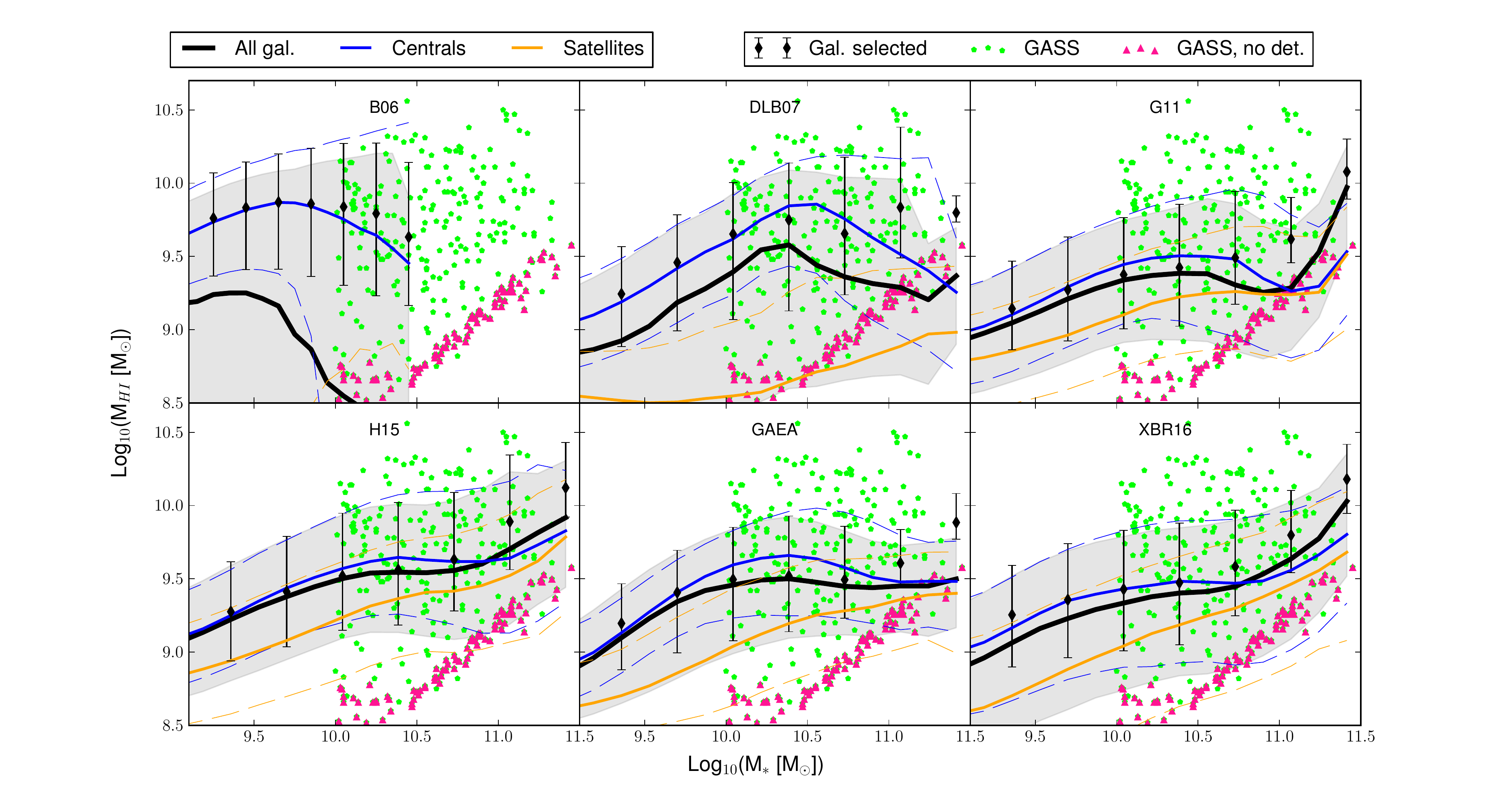} 
    \caption{The predicted relation between HI mass and galaxy stellar mass,
      compared to results from the GALEX Arecibo SDSS Survey (symbols). Lines
      of different colours are used for central galaxies (blue) and satellites
      (orange), while the black lines and shaded regions correspond to the
      entire galaxy population. Solid lines correspond to the median of the
      distributions, while dashed lines (and the shaded region) mark the 
      1$\sigma$ distribution. The black diamonds with error-bars show
      the distribution of all model galaxies when using the same detection
      threshold of the data. Symbols of different colours are used for 
      detections (green pentagons) and non detections (upper limits, purple
      triangles).} 
    \label{fig:HI_Ms}
\end{figure*}

Fig.~\ref{fig:HI_Ms} shows the predicted relation between the HI mass and the
galaxy stellar mass for all models used in our study. Solid lines show the
median of the distributions, while dashed lines (and the shaded region that
corresponds to the black line) mark the 1$\sigma$ spread. Lines of
different colours correspond to centrals (blue) and satellite galaxies
(orange). Symbols correspond to observational data from the GALEX Arecibo SDSS
Survey \citep{catinella2013}, divided in detected HI (green pentagons) and 
non detections, and thus upper limits (purple triangles).

Given the large spread of the observational data and the detection limits, it
is hard to use these data to put strong constraints on the models. All models
cover the region sampled by the observational data but for the B06 model, that
predicts no HI in galaxies more massive than $\sim 10^{10}\,{\rm M}_{\sun}$
likely because of too efficient AGN (radio-mode) feedback.  We accounted for
the same detection limit of observations (for $M_*>10^{10}\;M_{\sun}$, only
considered galaxies with M$_{HI}$ larger than the maximum no detection in the
observed sample). Results are shown in this case as black diamonds with 1$\sigma$
error-bars: this narrows the distributions of model galaxies, particularly for
the G11, GAEA, and XBR16 models. Overall, the agreement between model
predictions and observations is good, but some models (e.g. G11, GAEA and
XBR16) exhibit a deficit of HI-rich galaxies at any given stellar mass. This remains
statistically significant even when applying a detection threshold similar to
that of the survey used as a comparison.

As expected, central galaxies always have larger HI masses than satellite
galaxies. The difference is largest for the B06 and DLB07 models that adopt an
instantaneous stripping of the hot gas reservoir associated with infalling
satellites. The G11 and H15 models relax this assumption, which brings the
median HI mass of satellite galaxies closer to that of central
galaxies. Interestingly, the difference between these medians in the GAEA and
XBR16 models, that also assume an instantaneous stripping of the hot gas, is
only slightly larger than that found in G11 and H15. As discussed in
\citet{hirschmann2015}, this is driven by significantly larger amounts of cold
gas at higher redshift and a reduced reheating rate from stellar feedback.
Interestingly, at the most massive end, the HI content of satellite galaxies is
comparable to that of centrals in all models but B06 and DLB07.

\begin{figure*}
  \includegraphics[trim=0.5cm 0.8cm 1cm 0.1cm, clip, width =
    \textwidth]{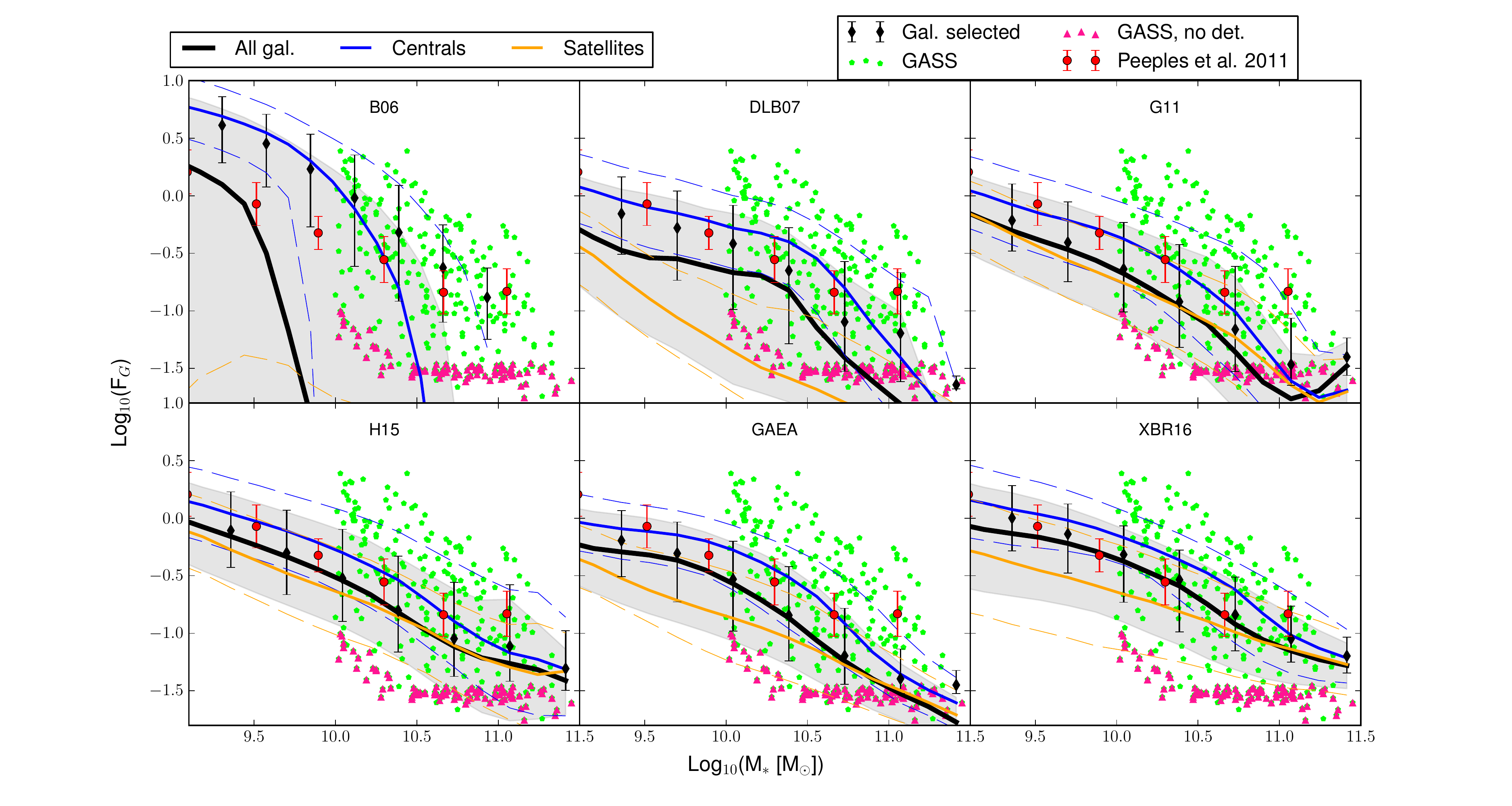} 
  \caption{The gas fraction $F_G=M_{cold}/M_*$ of model galaxies as a function 
  of their stellar mass, compared to observational measurements by
    \citet[][green pentagons and pink triangles]{catinella2013}, and \citet[][red 
    circles with error
      bars]{peeples2011}. Lines (solid and dashed), shaded region and diamonds with
      error-bars have the same
    meaning as in Fig.~\ref{fig:HI_Ms}.}
  \label{fig:Fg_Ms}
\end{figure*}

In Fig.~\ref{fig:Fg_Ms}, we show the cold gas fraction $F_g = M_{cold}/M_{*}$
 as a function of the galaxy stellar mass for all models considered
in our study. Symbols show observational measurements by \citet[][GASS
survey, pentagons and triangles]{catinella2013} and \citet[][red points with error
bars]{peeples2011}. The latter are a collection of previous HI measurements
from \citet{mcgaugh2005}, \citet{west2009, west2010} and \citet{leroy2008}. 
To convert the measurements in cold gas fraction, we have assumed HI 
represents a constant fraction ($\sim$71 per cent) of the total cold gas available.

As discussed above, the B06 model
clearly under-predicts the gaseous content of galaxies of intermediate to high
mass. In addition, virtually all gas in this model is associated with central
galaxies. Comparing Fig.~\ref{fig:Fg_Ms} and Fig.~\ref{fig:HI_Ms} with the predicted
HI mass function for this model (Fig.~\ref{fig:MF_HI}), we can see that the
excess of HI rich galaxies is driven by the high gas fractions of galaxies
with low stellar masses. As noted above, even accounting for the same selection
limits of the observations (black diamonds and error-bars), 
only some of the models predict gas fractions as 
large as those observed (DLB07, H15, XBR16). 
Finally, both the DLB07 and the 
GAEA models under-predict the gas fractions estimated for galaxies with the 
largest stellar mass. When applying the same selection of the data, however, 
model predictions appear consistent with observations.
It should be noted that
observational measurements are sparse and likely incomplete in this mass
regime, thus further measurements are necessary to constrain this relation
at the highest stellar masses.

\section{Two point projected correlation function of HI selected galaxies}
\label{sec:2PCF_theory_models}

In this section we will analyze the clustering properties of model galaxies,
selecting them in HI bins. We use, as observational reference, the work by
\citet[][P13 hereafter]{papastergis2013}. They used 6,123 HI-selected galaxies
from the ALFALFA survey, covering a contiguous rectangular sky region of $\sim
1,700$ deg$^2$ in the redshift range $z\sim 0.0023-0.05$.  They also used an
optical sample of 18,516 galaxies in the same volume using the SDSS DR7
\citep{abazajian2009}, and applying a magnitude cut of $M_r<-17$.  As expected
and shown in previous studies \citep[e.g.][]{catinella2010,huang2012}, 
the most luminous galaxies tend to be HI poor,
while HI rich galaxies tend to be the dominant population among galaxies that
populate the blue cloud.

Below, we will compare our model predictions with the estimated projected
correlation functions for the following HI mass bins:
$\log_{10}(M_{HI}[M_{\sun}])\in[8.5;\;9.5]$, $[9.5;\;10.0]$ and
$[10.0;\;10.5]$. 
Lower mass bins were considered in P13, but these
fall below the resolution limit of our models. The lowest bin  considered 
here is already partially below the completeness limit. We will account
for this in the following.

P13 found little evidence 
of dependence of the clustering signal on the HI content, with some 
uncertainties for the lowest HI
mass bin. For galaxies in this bin, a lower clustering signal is
measured with respect to HI richer galaxies, but P13 argue this is not
statistically significant due to the lower volume sampled by galaxies in this
HI bin. 

\subsection{The projected correlation function}
We compute the two-point correlation function (2PCF) for all models used in our
study taking advantage of the mock light-cones described above in
Sec.~\ref{sec:lightcone}. To compute the predicted 2PCF, we use the
\citet{landy1993} estimator:
$$\xi(r)=\frac{DD(r)+RR(r)-2DR(r)}{RR(r)}$$ where $DD(r)$, $RR(r)$ and $DR(r)$
represent the galaxy-galaxy, random-random, and galaxy-random number of galaxy
pairs separated by a distance $r$. 
In observations, the physical separation $r$ is not directly available, and
the observables are the position on the sky and the recessional velocity.
Using our mock light-cones (Sec.~\ref{sec:lightcone}) we mimic the data and 
carry out all calculations in redshift space. We measure the separation among 
two galaxies 
as $s=\sqrt{(v_1^2+v_2^2-2v_1v_2\cos{\theta})}/H_0$, where $v_1$ and $v_2$ are 
the recessional velocities of the galaxies (km s$^{-1}$), $\theta$ is the 
angle between them in the sky, and $H_0$ is the Hubble constant. 
Hence we calculate the correlation function using the separation
along the line of sight ($\pi=|v_1-v_2|/H_0$) and on the sky plane 
($\sigma=\sqrt{\pi^2-s^2}$) to obtain $\xi(\pi,\sigma)$.
The projected correlation function used in the following discussion 
corresponds to $$w(\sigma)=2\int^{\infty}_{0} \xi(\pi,\sigma)d\pi.$$

P13 had a non-uniform radial selection function for their
sample and accounted for it in their random sample. 
We do not attempt to mimic the selection function of the data, and simply use 
for the random catalogue the same smooth redshift distribution of the selected 
model galaxies.

\subsection{Model predictions}
\label{sec:results_HI_clustering}

\begin{figure*}
  \includegraphics[trim=2.5cm 0.5cm 0 .0cm, clip, width =  0.9\paperwidth]{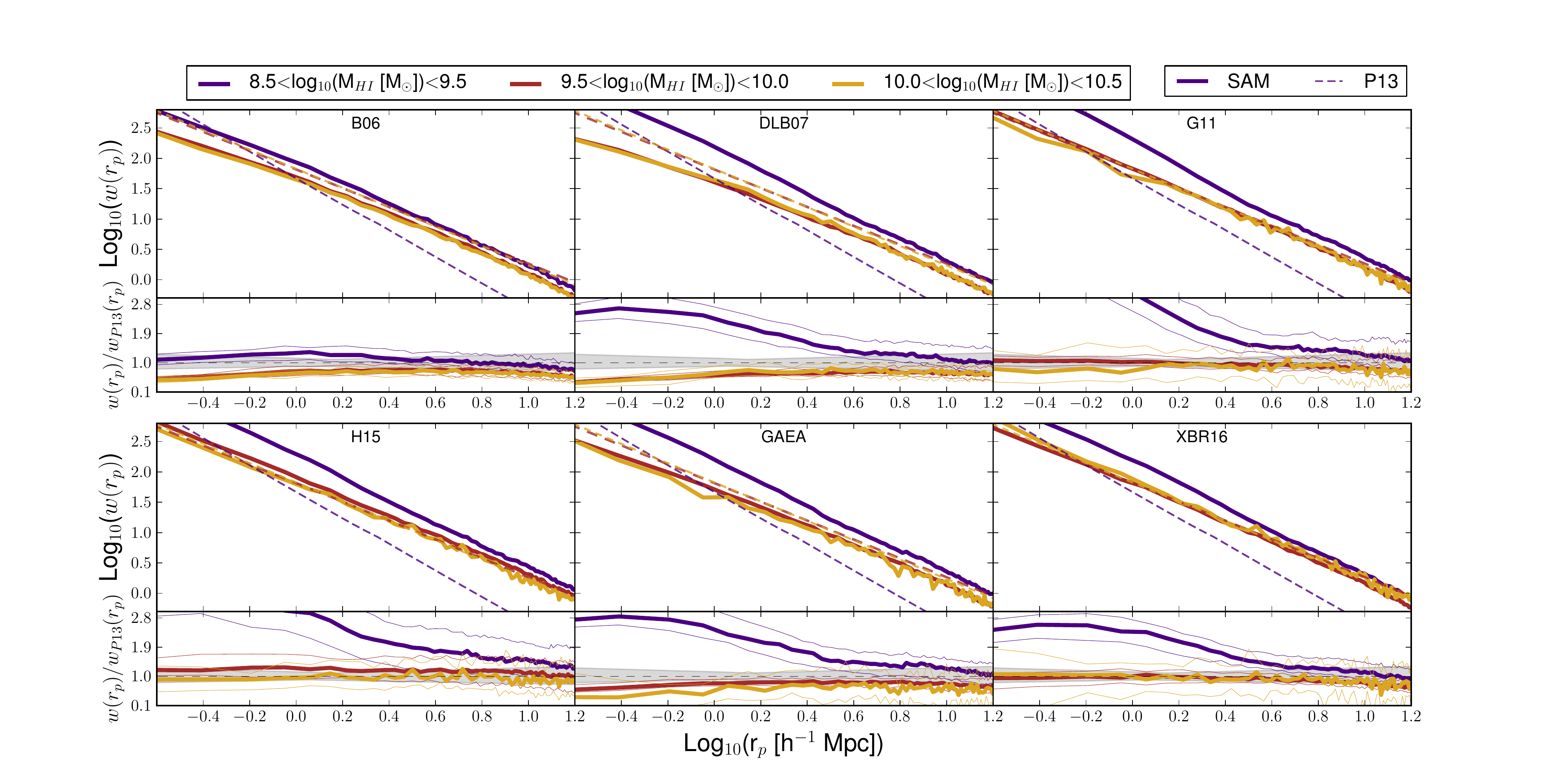}
  \caption{The two-point projected correlation function for different bins in
    HI mass (solid thick lines of different colours). Each panel corresponds to
    a different model, and the lower sub-panels show the ratio between the
    model predictions and the corresponding observational estimates (dashed
    lines) by P13 for the medium HI mass bin. As discussed in the text, this is
    equal to the one of  
    the highest  HI mass bin and more reliable than the measurements
    corresponding to the lowest HI mass bin. 
    The 1$\sigma$ scatter is shown in the lower
    sub-panels using thin lines for models and a shaded region for the
    data.}
  \label{fig:corr_HI_bin}
\end{figure*}

Fig.~\ref{fig:corr_HI_bin} shows the 2PCF for all models considered in this
study. 
P13 found that the middle and highest HI mass bins 
($M_{HI} \in [10^{9.5};\;10^{10}]$ and $[10^{10};\;10^{10.5}]\;M_{\sun}$)
have almost the same 2PCF, 
while they considered the measurements obtained for the lowest HI bin 
($M_{HI} \in [10^{8.5};\;10^{9.5}]\;M_{\sun}$) not 
reliable due to the small sampling volume. To better show the differences 
between model predictions and observational measurements, we show in the bottom 
sub-panels model results in each HI mass bin divided by the observational 
measurement corresponding to the bin $M_{HI} \in [10^{9.5};\;10^{10}]\;M_{\sun}$.

For the lowest HI mass bin ($[10^{8.5};\;10^{9.5}]\;M_{\sun}$), a problem is immediately evident: P13 found
these galaxies to be less clustered than their HI richer counterparts, but 
argued that this might be due to finite volume effects. All models considered 
in this study, in contrast, predict a higher clustering signal for galaxies 
in this bin. The B06 model is the closest to the observational measurements, 
but still standing a little above them.
All other models largely over-predict the measured clustering signal, particularly at small
scales, with G11 and H15 deviating most from the data. As we will see in the
following, this HI mass bin is sensitive to various physical prescriptions and
to numerical resolution. In addition, it is dominated by satellite galaxies and
therefore strongly dependent on the adopted treatment for satellite evolution.
 We verified how the incompleteness of the model sample influences  
the final 2PCF of this bin for the G11, H15 and XBR16 models using the MS-II:
we find that the predicted clustering signal is lower than that found for the MS, 
but still a factor $\sim 2$ larger than that measured.

For HI richer galaxies ($[10^{9.5};\;10^{10}]$ and $[10^{10};\;10^{10.5}]\;M_{\sun}$), 
B06, DLB07 and GAEA systematically under-predict the
clustering signal, while G11, H15 and XBR16 are in good agreement with
observational measurements in the corresponding mass bins.

The relatively noisy behaviour of the 2PCF in the highest HI bin is due to 
the small number of galaxies with such large HI masses.

\begin{figure}
  \includegraphics[trim=0.5cm 0cm 1cm 1cm, clip, width =
    \columnwidth]{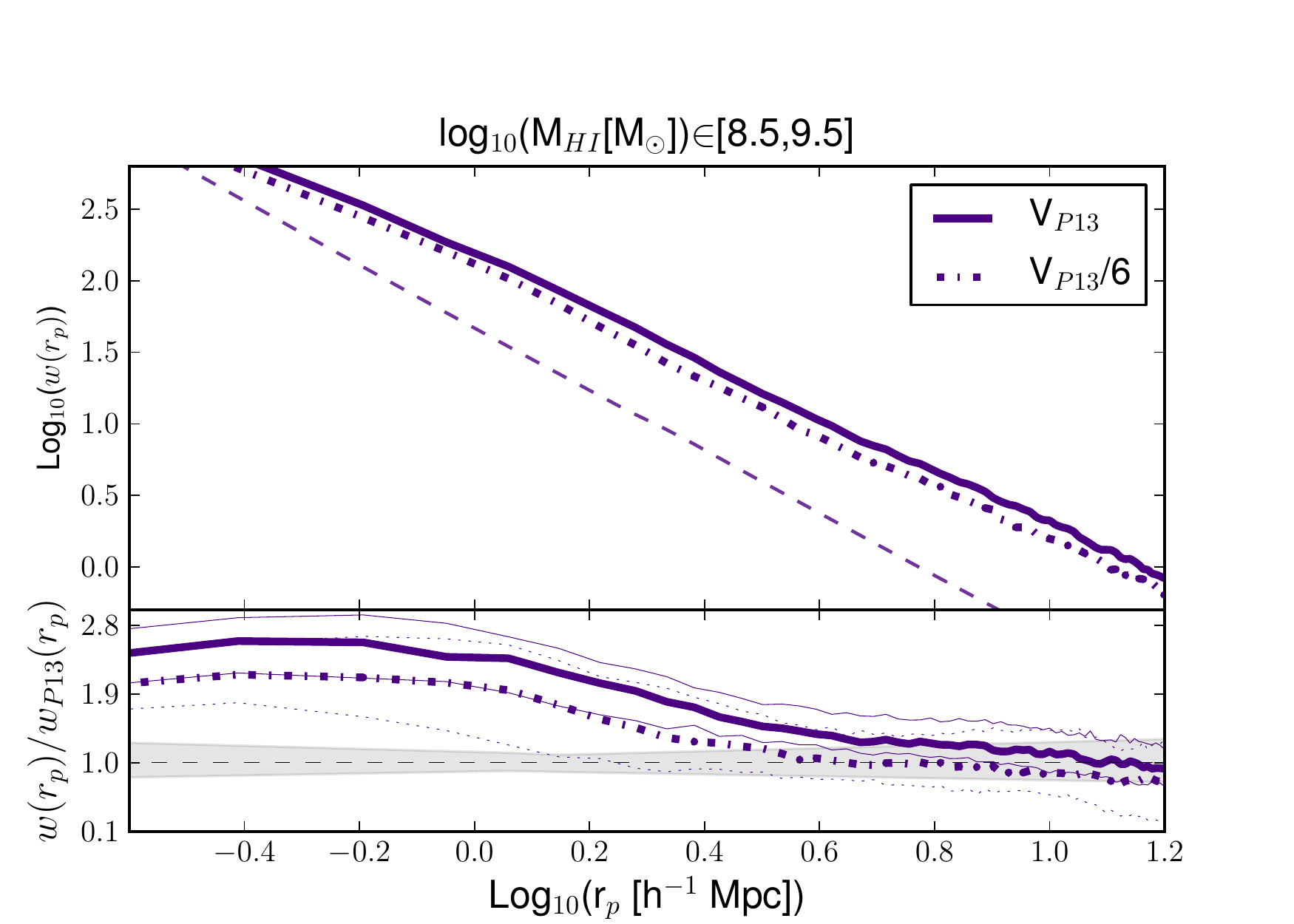} 
  \caption{The 2PCF for the lowest HI mass bin for the XBR16 model. The solid
    line corresponds to model predictions for the full volume considered
    (i.e. this line corresponds to that shown in
    Fig.~\ref{fig:corr_HI_bin}). The dot-dashed line corresponds to the results
    obtained when considering only one sixth of the volume, as in
    the observational sample by P13. In the lower sub-panel, we show the ratio
    between model predictions and the observational estimate of the medium HI 
    bin. Shaded regions and thin lines show the 1$\sigma$ scatter.}
  \label{fig:corr_HI_bin_smallV}
\end{figure}

As noted above, P13 argue that for the lowest HI mass bin, results are affected
by smaller sampling volume. We can test the influence of a small sampling
volume using our mocks. We show results of this test in
Fig.~\ref{fig:corr_HI_bin_smallV}. In this case, we use only the XBR16
model (results are similar for the other models) and show the 2PCF 
only for the lowest HI mass bin. The corresponding model 
predictions obtained by using only
one sixth of the volume are shown as a dotted-dashed line, and are not
statistically different from those obtained using the entire volume (solid
line). 

\subsection{Halo Occupation Distribution}
\label{sec:results_HI_hod}

The results in the previous section can be understood by considering
the number occupation of haloes of different mass by galaxies with different HI
content, i.e. what is typically referred to as halo occupation
distribution (HOD). The HOD gives information about the distribution
of galaxies in dark matter haloes of different masses, with the possibility to 
distinguish between centrals and satellites. It can be used to interpret 
the 2PCF at the scales of the halo dimensions. 
 For this analysis, we use the data from the z=0 snapshots
to have a larger statistics.

\begin{figure*}
  \includegraphics[trim=30 3cm 0.5cm 3cm, clip, width = 0.6
    \paperwidth]{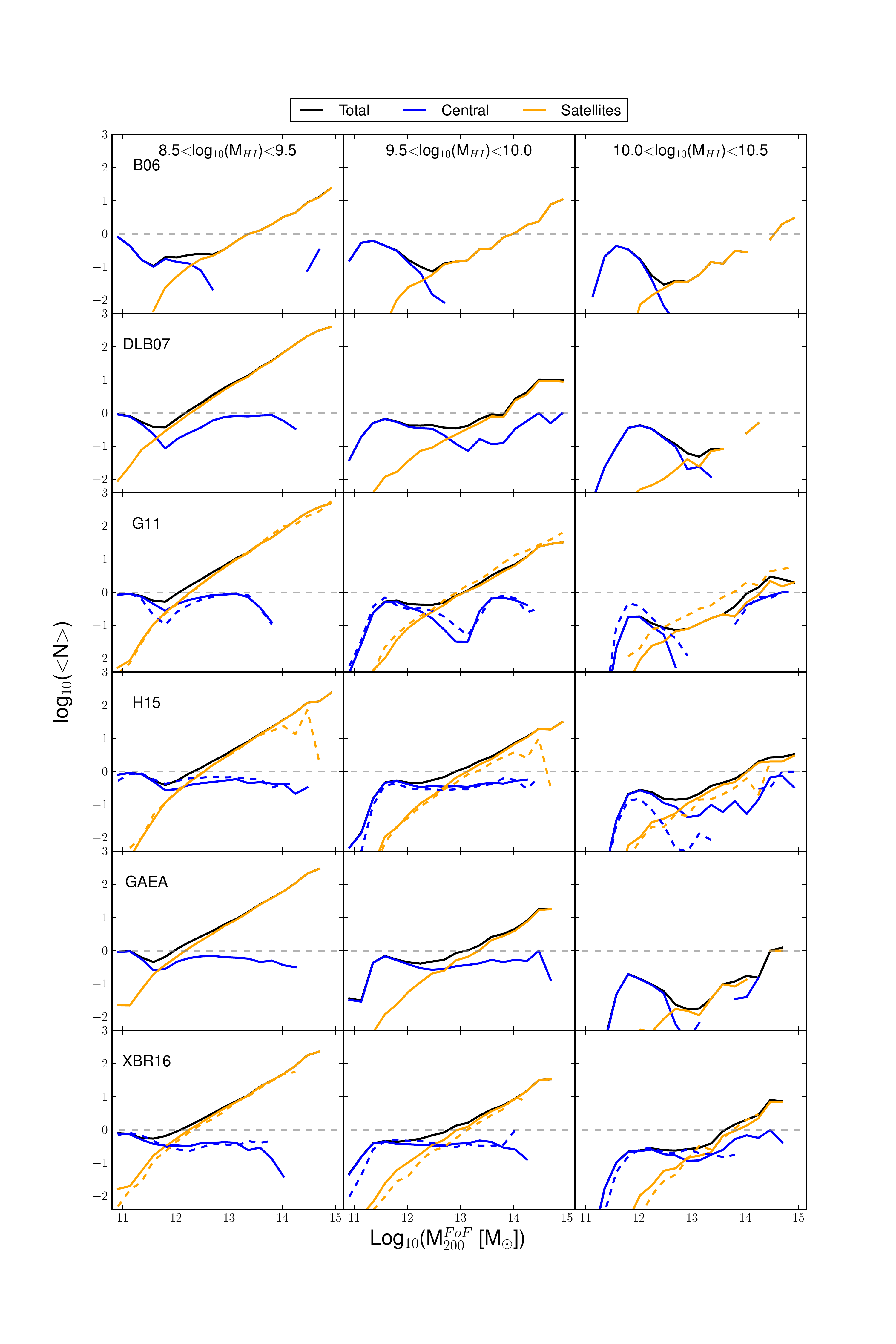} 
  \caption{Average number of galaxies with stellar mass larger than
    $M_*>10^{9}M_{\odot}$ and in different HI mass bins (different
    columns) for haloes of mass $M_{200}^{FoF}$. Different rows correspond to the different models considered in
    our study. The colour coding is the same as in Fig.~\ref{fig:HI_Ms}: black
    for all model galaxies, blue for centrals and orange for satellites. In the rows 
    corresponding to the G11, H15 and the XBR16 models, the dashed lines 
    correspond to results based on the MS-II simulation. }
  \label{fig:hod_hi_in_all}
\end{figure*}

Results from all models used in this study are shown in
Fig.~\ref{fig:hod_hi_in_all}. Each panel shows the average number of galaxies
with stellar mass larger than $M_*>10^{9}M_{\odot}$ (approximately
corresponding to the resolution limit of the MS) and with
different HI mass (different columns) that reside in FoF haloes of mass $M_{200}^{FoF}$. 
Blue and orange lines correspond to
central and satellite galaxies respectively, while black solid lines show the
total. We checked the convergence of the results obtained for low mass FoF haloes
by considering the G11, H15 and the XBR16 model on the MS-II (results are shown as 
dashed lines in the corresponding rows).
The HOD of the central galaxies has a Gaussian shape in B06 
\citep[as already noted in ][]{kim2011}, with basically no central galaxies for
halo masses larger than  $\sim 
10^{12.5}\;M_{\sun}h^{-1}$. In the other models the distributions corresponding
to central galaxies are generally broader and extend to significantly larger
halo masses than in B06. As noted above, HI in central galaxies of massive
haloes is likely suppressed in B06 by efficient radio-mode feedback. 

\begin{figure*}
  \includegraphics[trim=3cm 1.5cm 3.0cm 2.0cm, clip, width = 0.8 \paperwidth]{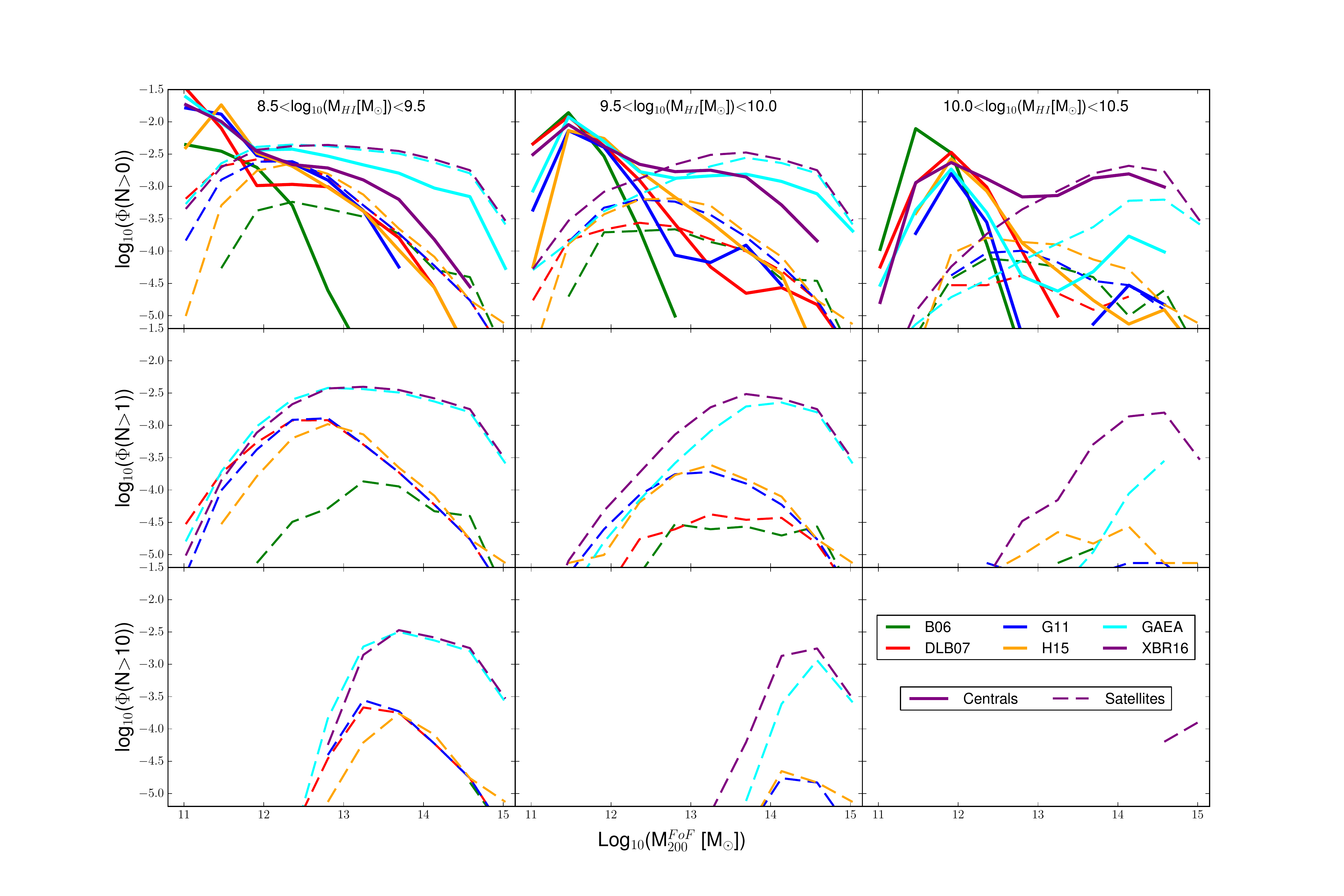}
  \caption{Mass function of FoF haloes hosting at least one (top panel),
	more than one (middle panel), and more than ten (bottom panel)
	galaxies with HI mass in different bins (different columns).
	Solid lines are used for central galaxies while long-dashed lines 
	correspond to satellite galaxies.
	}
  \label{fig:MF_fof}
\end{figure*}

In Fig.~\ref{fig:MF_fof} we show the mass function of FoF haloes 
hosting N galaxies with HI mass in different bins (different columns), 
namely the volume density (per $M_{200}^{FoF}$ bin) of haloes hosting at least 
N centrals/satellites with a selected HI content. 
Different rows correspond to N>0, N>1 and N>10. 
Galaxies are divided
in centrals (solid lines) and satellites (dashed lines). Note that the 
FoF mass function for the central galaxies is shown only in the N>0 row, as each FoF 
contains only one central galaxy by construction. 
Only  a few haloes have large number of satellites containing large
amounts of HI. The GAEA and XBR16 models in particular, have significantly
larger number of satellite galaxies with modest to significant HI content with
respect to the other models. For all other models, the distributions peak at
lower halo masses. It is interesting that GAEA and XBR16 predict a significant
contribution from the HI gas rich satellite population albeit assuming an
instantaneous stripping of the hot gas associated with infalling galaxies. As
noted above and in \citet{hirschmann2015}, this is driven by the modified
stellar feedback scheme. For the intermediate and high HI mass bins 
considered ($[10^{9.5};\;10^{10.5}]\;M_{\sun}$),
there are less haloes hosting large numbers of galaxies in the GAEA model than
in XBR16. We remind that these two models are based on the same physical
parametrizations and differ only for the explicit modelling of the partition of
cold gas in atomic and molecular gas, the division of the disk in annuli, and 
on the explicit dependence of star
formation on the molecular gas content in XBR16.

In the top row of Fig.~\ref{fig:MF_fof} , haloes hosting only one central galaxy dominate the 
low $M_{200}^{FoF}$ end, and outnumber
the FoF with HI-selected satellites up to $M_{200}^{FoF}\sim 10^{12}\;M_{\sun}h^{-1}$.
Above this limit, HI rich satellites become the main contributors.

In Fig.~\ref{fig:hod_hi_in_all}, in the lowest HI mass bin ($[10^{8.5};\;10^{9.5}]\;M_{\odot}$), satellites
outnumber centrals at $M_{200}^{FoF}\sim 10^{12}\;M_{\sun}h^{-1}$ in the
B06 and H15 models. In DLB07, G11, GAEA and XBR16, the number of satellites
becomes larger than the corresponding number of centrals at
$M_{200}^{FoF}\sim 10^{11.5}\;M_{\sun}h^{-1}$. All models have more than 10
galaxies per FoF only above $M_{200}^{FoF}\sim 10^{13}\;M_{\sun}h^{-1}$ (B06
only at $\sim 10^{14}$). 
The volume density of such haloes is significant only for the GAEA and 
XBR16 models.
B06 has the lowest number of haloes with more than one satellite in all the 
HI mass bins, but the difference in the lowest bin is more relevant, in particular
compared to the DLB07 model, which instead is aligned with G11 and H15 both 
in the HOD and in the mass function. This results in different 2PCFs for B06 and the
other models.

For the middle HI mass bin ($[10^{9.5};\;10^{10}]\;M_{\odot}$), we always find less
than 10 galaxies per halo in the B06 and DLB07 models. 
In all other models, haloes with mass larger than $M_{200}\sim
10^{14}\;M_{\odot}h^{-1}$ host more than 10 satellite galaxies.
However, as for the lowest HI mass bin, the number of these haloes is
large only for the GAEA and XBR16 models.
In general B06 and DLB07 have less satellite galaxies than the other models. 
This can explain the underestimation of the 2PCF: a lower number of satellite 
galaxies lowers the correlation signal.
We remind that these models are characterized by a simple treatment for 
satellites and  gas stripping (see Sec.~\ref{sec:results_HI_satellites} for 
details on satellite evolution), that leads to the well known problem of too 
many passive galaxies \citep{weinmann2006,wang_li2007,fontanot2009}. This simplified treatment is, however, assumed also
in the GAEA and XBR16 models, in which the effect of the instantaneous
stripping of gas is mitigated by a different treatment of stellar feedback.

There are generally very few galaxies in the highest HI mass bin
($[10^{10};\;10^{10.5}]\;M_{\odot}$) considered. In particular, the B06, DLB07 and
GAEA models always have only one galaxy per halo below halo masses $M_{200}\sim
10^{14.5}\;M_{\odot}h^{-1}$. For the other models, the number of galaxies
becomes larger than one at halo masses larger than $M_{200}\sim
10^{13.7}\;M_{\odot}h^{-1}$.
The only model with numerous satellites is XBR16. 
The difference in number of satellite rich haloes does not determine an 
appreciable difference between the XBR16, G11 and H15 2PCFs. The smaller
numbers of satellite rich halos in the GAEA model result in an underestimated 
2PCF. The difference with respect to the XBR16 model can again be ascribed to
the approach adopted, based on dividing the star forming disk in annuli. As
commented above, this results in a lower molecular fraction to total cold gas
with respect to the GAEA model, and keeps the star formation ongoing in the
central regions of the disk for longer times. This implies that satellite
galaxies in XBR16 have more HI left than their counterparts in the GAEA model.

\section{The role of satellite galaxies}
\label{sec:results_HI_satellites}
The results discussed above suggest that satellite galaxies play an important
role in the disagreement found between model predictions and observational
measurements. 

\begin{figure*}
  \includegraphics[trim=2.5cm 0.5cm 0 .0cm, clip, width = 0.9
    \paperwidth]{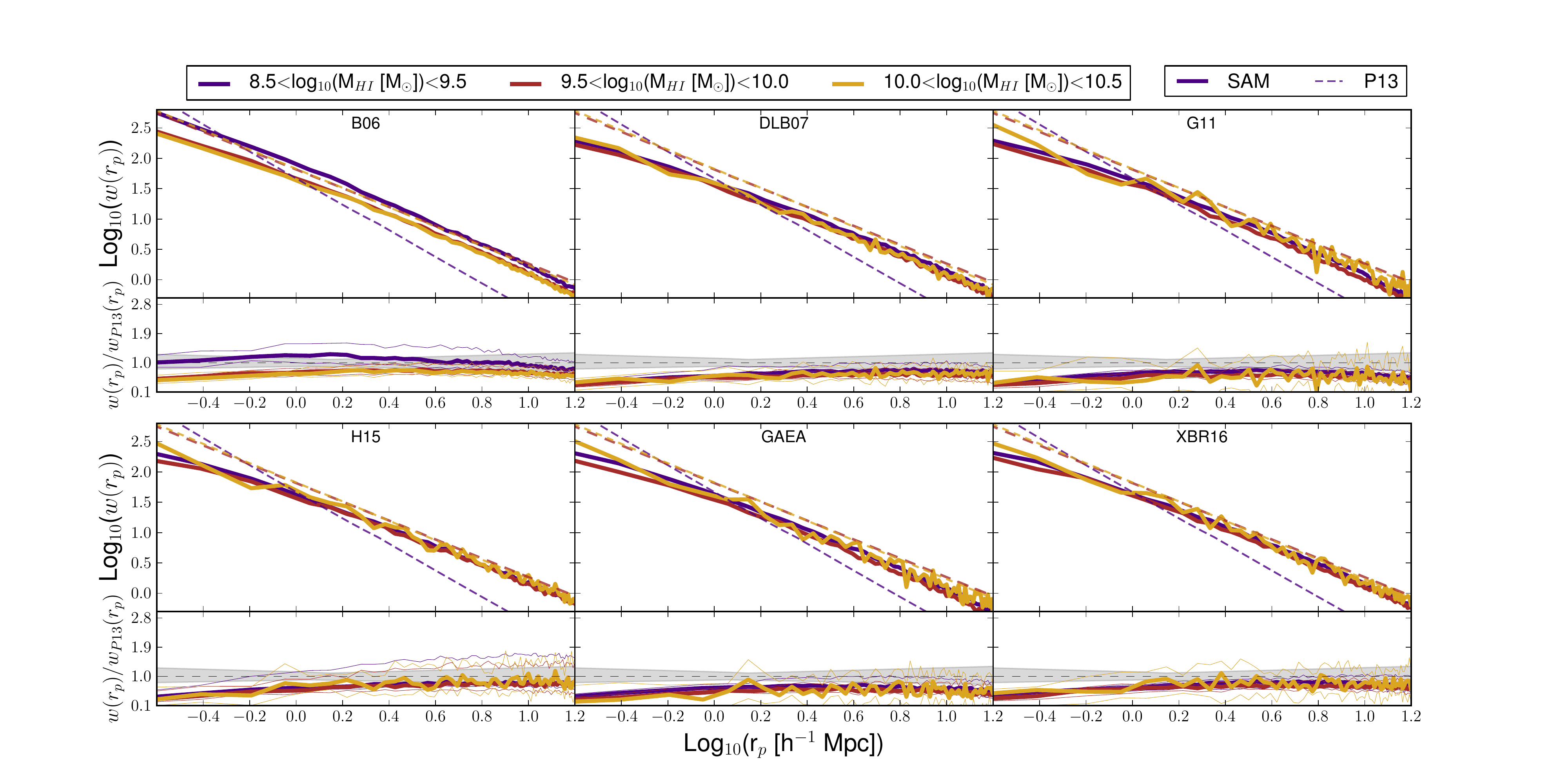} 
  \caption{As in Fig.~\ref{fig:corr_HI_bin}, but considering only
    central galaxies for all models. }
  \label{fig:corr_HI_bin_no_sat}
\end{figure*}

As discussed in Section~\ref{sec:SIMSAM}, all models used in our study are
built using subhalo based merger trees extracted from the Millennium
Simulation. Dark matter haloes are subject to significant stripping after being
accreted on larger systems \citep[e.g.][]{delucia04b,gao04}. At the resolution
of the MS, substructures fall below the resolution limit
when at distances from halo centre that are typically significantly larger than
the separation from which galaxy mergers are expected to occur. All models used
in our study assume that galaxies in disrupted subhalos survive as `orphan'
galaxies. While the specific treatment depends on the model (e.g. G11 and H15
include an explicit treatment for tidal disruption), they all assign to
these galaxies a residual merger time that generally depends on the initial
orbit and on the mass-ratio between the infalling system and the accreting
one. 

 When the infalling process begins satellites undergo tidal 
processes that strip away a part or all their hot gas halo. In B06, DLB07,
GAEA and XBR16 this process is instantaneous, and all the hot gas is stripped
away (in B06, all the gas outside the dynamical radius of the halo) at the 
infall time (when the galaxy becomes a satellite). In G11 and H15, the 
stripping is gradual, and the hot gas that remains associated with satellite 
galaxies can cool providing fresh material for star formation.

The effect of satellite galaxies on the predicted correlation function of HI
selected galaxies is shown in Fig.~\ref{fig:corr_HI_bin_no_sat}.  In this case,
we are considering only central galaxies, i.e. we are excluding from
galaxy catalogues both `orphan' galaxies and satellite galaxies associated with
distinct dark matter substructures. 
In all models, the clustering signal becomes weaker, with a shift  
dependent on the number of the satellites in each HI mass bin.
For the B06 model, the 2PCF remains the same as if satellites are included. 
This is expected as Fig.~\ref{fig:HI_Ms} shows that satellite
galaxies in this model are typically HI poor.
Also for the DLB07 model, small differences are found between the clustering 
signal predicted including and excluding satellite galaxies in the medium-high
HI bins ($[10^{9.5};\;10^{10.5}]\;M_{\sun}$).  
Again, this is expected because satellite galaxies are HI poor in this model,
although the effect is not as strong as for the B06 model.

In Tab.~\ref{tab:percentages} the fraction of satellites in each HI mass
bin is listed for each model considered. Satellites account for 
$39-46\%$ of the total, depending on the model. 
The majority are in the lowest HI bin ($[10^{8.5};\;10^{9.5}]\;M_{\sun}$) 
considered in this analysis, with the 
exception of the B06 model where satellites have typically lower HI masses.

\begin{table*}
\begin{center}
\caption{Fraction of central and satellite galaxies in different HI mass bins, 
	for all models considered in this study ($M_*>10^9M_{\sun}$).}
\label{tab:percentages}
\begin{tabular}{c|ccccc|}
Model & $\%$ of centrals/ & \multicolumn{4}{c}{$\%$ of central/satellites with $M_{HI}[M_{\sun}]\in$}\\
      & satellites&$[0;\;10^{8.5}]$ & $[10^{8.5};\;10^{9.5}]$ & $[10^{9.5};\;10^{10}]$ & $[10^{10};\;10^{10.5}]$\\\hline
      
B06	& 54.3 - 45.7	& 7.8 - 39.5	& 5.6 - 3.1	& 20.4 - 1.4	& 16.8 - 0.6\\
DLB07	& 54.4 - 45.6	& 1.5 - 19.8	& 33.0 - 24.6	& 15.6 - 1.2	& 4.0 - 0.1\\
G11	& 53.3 - 46.7	& 1.9 - 3.7	& 34.0 - 37.6	& 14.1 - 4.8	& 3.2 - 0.6\\
H15	& 61.4 - 38.6	& 0.9 - 2.3	& 33.7 - 28.8	& 20.2 - 6.3	& 5.7 - 1.0\\
GAEA	& 59.3 - 40.7	& 1.0 - 7.2	& 39.7 - 30.9	& 16.8 - 2.6	& 1.8 - 0.1\\
XBR16	& 57.0 - 43.0	& 4.7 - 12.7	& 35.0 - 25.8	& 15.0 - 4.1	& 2.3 - 0.4\\
\end{tabular}
\end{center}
\end{table*}

\begin{figure*}
  \includegraphics[trim=1.0cm 0.5cm 0 .0cm, clip, width = 0.8
    \paperwidth]{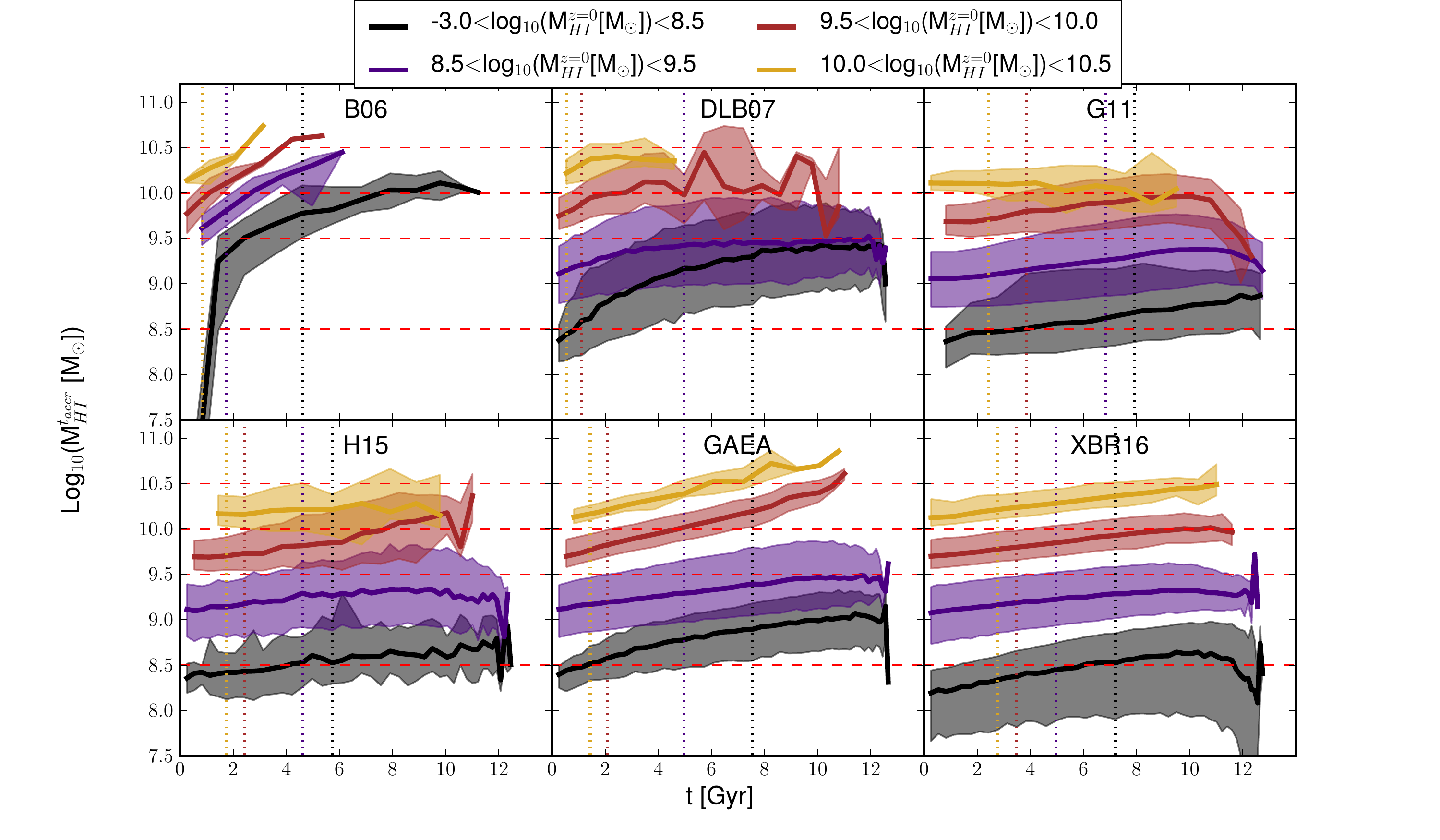} 
  \caption{The HI content of satellites at the time of accretion vs
	   the look-back time of accretion. 
	   Galaxies are divided according to their final HI content, with
	   different colours representing different HI mass bins. The solid 
	   lines correspond to the median of the 
	   distributions, while the shaded areas show the one sigma scatter.
	   The vertical dotted lines are the median times of accretion of 
	   model galaxies in each HI bin (colour coded as above). 
	   The red horizontal dashed lines correspond to the limits of the HI mass bins 
	   considered, and are plotted as a reference.}
  \label{fig:med_HIz_t}
\end{figure*}

Taking advantage of model results, we can also quantify the evolution of the 
HI content in satellite galaxies. To this aim, we have selected all 
satellite galaxies at $z=0$ and followed back in time their main progenitors
until they become central galaxies. In Fig.~\ref{fig:med_HIz_t}, we show the HI
mass that satellite galaxies have at the last time they are centrals, as a 
function of lookback time. Galaxies are split in different bins according to 
the HI mass at present. We plotted as vertical dotted lines the median times
of accretion for each HI bin. The figure shows that, for the two bins corresponding 
to the largest HI masses, accretion times tend to be lower than for galaxies with 
lower HI mass. In other words, the HI richest galaxies tend to be accreted 
later. This is particularly significant for the B06 and DLB07 models that are
characterized by the most rapid consumption of the cold gas in satellites. The
figure also shows that for some models (e.g. B06, DLB7, GAEA) the slope of the 
lines tend to be steeper than for the other models, indicating a more rapid 
depletion of the HI content of model galaxies. 

\begin{figure*}
  \includegraphics[trim=1.0cm 0.5cm 0 .0cm, clip, width = 0.9
    \paperwidth]{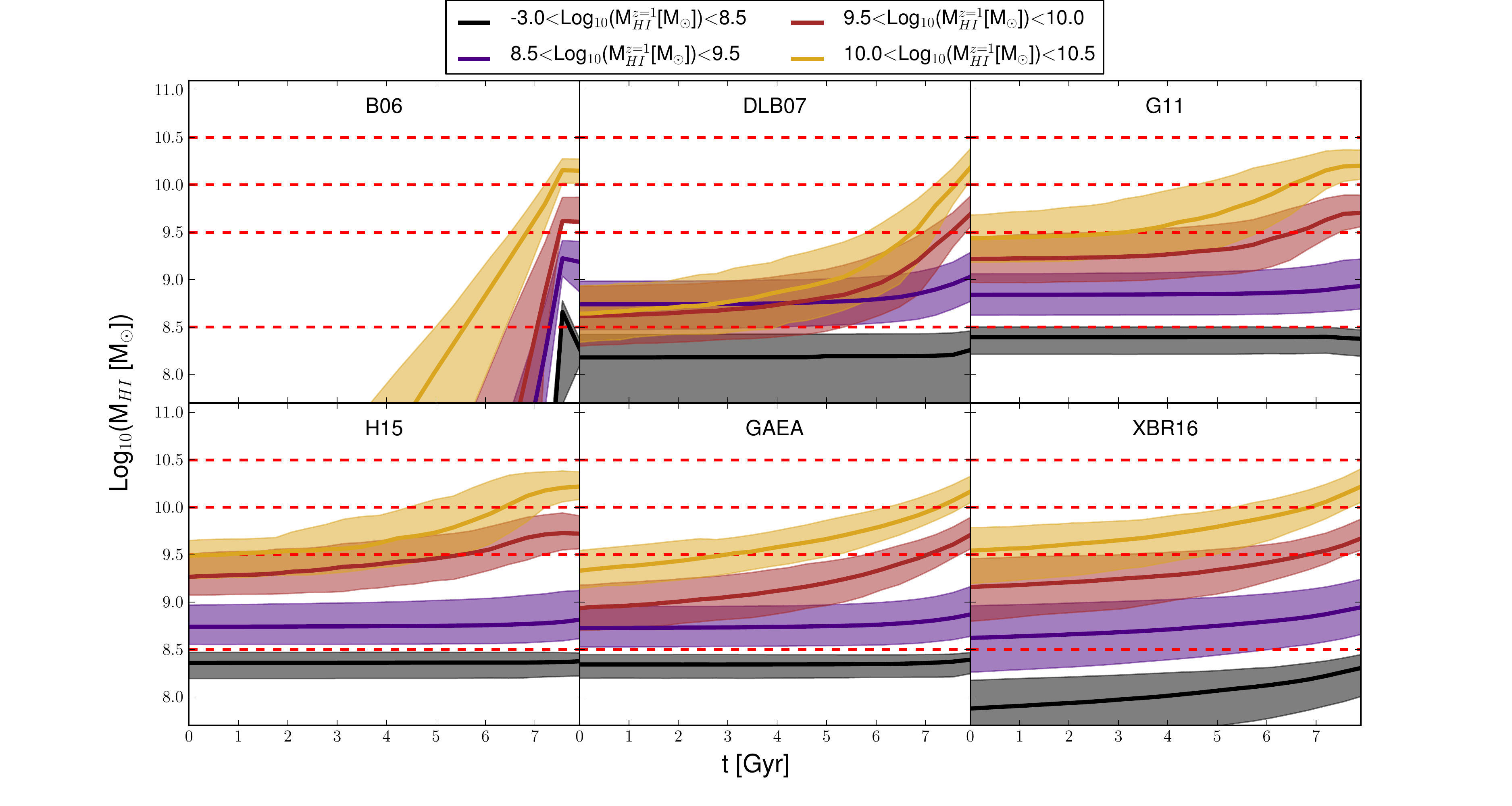} 
  \caption{The HI content of satellites accreted at redshift 1 vs
	   the look-back time. 
	   Galaxies are divided according to their initial HI content, with
	   different colours representing different HI mass bins at z=1. The solid 
	   lines correspond to the median of the 
	   distributions, while the shaded areas show the one sigma scatter.
	   The red dashed lines correspond to the limits of the HI mass bins 
	   considered, and are plotted as a reference.}
  \label{fig:med_HI_t_zfixed}
\end{figure*}

A more direct way to quantify gas depletion in satellite galaxies is to choose
a specific redshift of accretion and consider the average evolution of their 
HI content down to present time. Results of this analysis are shown in 
Fig.~\ref{fig:med_HI_t_zfixed} for satellites accreted at $z\sim 1$. Different 
colours correspond to different HI masses at the time of accretion. The B06 
model is characterized by the most rapid depletion rate. Also satellites in the
DLB07 model consume their gas rapidly but they tend to flatten when the average
HI content in satellites reaches a value $M_{HI}< 10^{8.7}\;M_{\sun}$. The figure
also clearly shows the different satellite treatment in G11 and H15 compared to 
GAEA and XBR16: the latter are characterized by a faster depletion rate soon 
after accretion. In DLB07, G11, H15, and GAEA the lines tend to flatten after
reaching some value. This is due to fact that these models assume a critical 
surface density of gas for star formation. In the XBR16 model, the flattening 
is less evident. This model does not assume an explicit threshold for star
formation but, as discussed above, evaluates the star formation rates in
different annuli after estimating the amount of molecular gas available using 
the \citep{blitz2006} empirical relation.

\section{Relations with the dark matter halo}
\label{sec:HI_DM}

In this section we analyze the relation between HI-selected galaxies
and the hosting dark matter haloes. We take advantage of our knowledge of the
halo mass to characterize the HI-$M_{halo}$ relation in 
Sec.~\ref{sec:HI_mhalo_max}, and we analyze the dependence of HI content on 
the dark matter halo spin in Sec.~\ref{sec:results_HI_spin}.

In the following, for satellite galaxies, we will consider the mass and spin of
the parent dark matter halo corresponding to the last time the galaxy was
central. 

\subsection{HI galaxy content and maximum halo mass}
\label{sec:HI_mhalo_max}
P13 used the measured 2PCF for the HI-selected galaxies to estimate the 
shape and the scatter of the HI-$M_{halo}$ relation. They took advantage 
of the Bolshoi dark matter only simulation \citep{klypin2011}, based on a WMAP7 
cosmology, and populated haloes and subhalos with HI through abundance matching.
In their analysis, they linked the current (z=0) HI content of each subhalo, 
$M_{HI}^{z=0}$, to the maximum value of the subhalo mass during its past 
evolution, $M_{200}^{max}$ (this corresponds with good approximation to the time just 
before a halo is accreted on a larger structure).
The implied assumption is that the HI attached to subhalos at their maximum 
mass does not change too much down to redshift 0. 
They note, but do not discuss further, that while this assumption can be 
valid for stellar masses, it is generally not a good one for the HI masses,
because of ram-pressure stripping and gas consumption due to quiescent 
star formation in satellite galaxies. 

\begin{figure*}
  \includegraphics[trim=2.5cm 0.5cm 2.5cm 0.5cm, clip, width = 0.8
    \paperwidth]{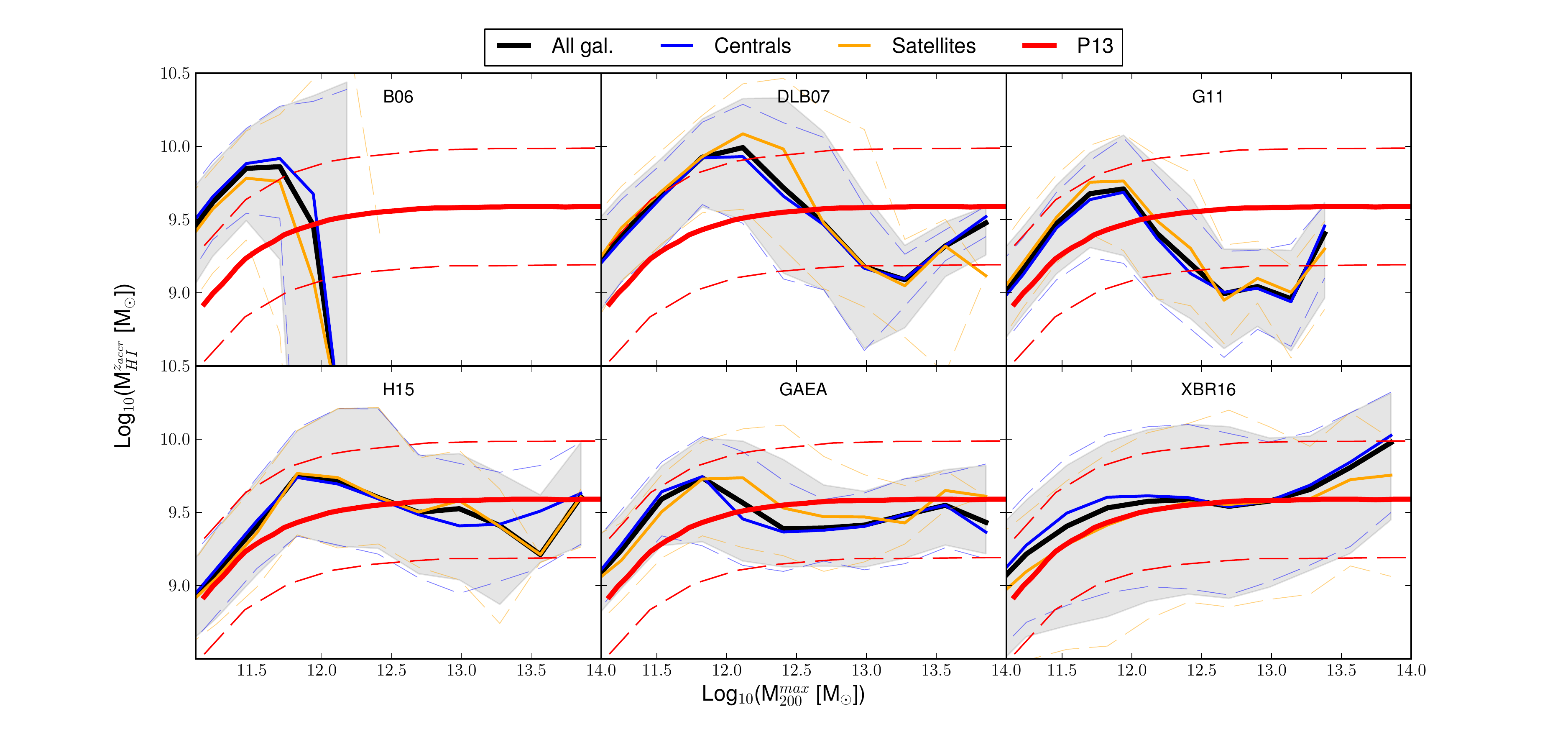} 
  \caption{The HI-halo mass relation for the models considered in this study,
    for central (blue), satellite (orange) and all (black) galaxies. The relation
    inferred by P13 is shown for comparison (red). The HI
    mass shown on the y axis is that measured at the time the parent halo mass peaks
    (i.e. just before infall).
    Solid lines show the median values, shaded region corresponds to the
    1$\sigma$ scatter for all galaxies, while dashed lines correspond to the
    1$\sigma$ scatter for centrals and satellites. }
  \label{fig:HI_Mh_ty0}
\end{figure*}

\begin{figure*}
  \includegraphics[trim=2.5cm 0.5cm 2.5cm 0.5cm, clip, width = 0.8
    \paperwidth]{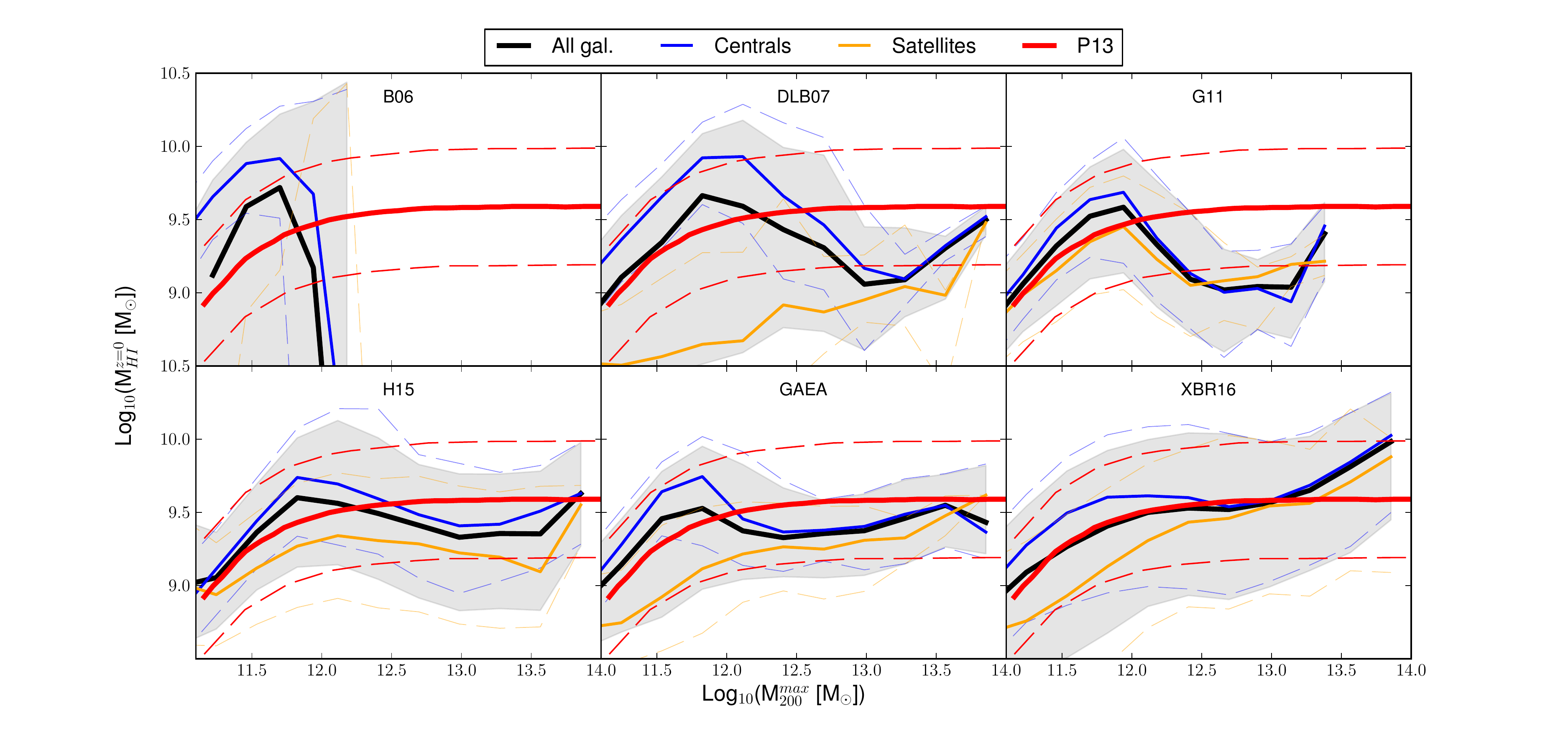} 
  \caption{As in Fig.~\ref{fig:HI_Mh_ty0}, but the HI
    mass shown on the y axis is that measured from the models at redshift 0.}
    
    \label{fig:HI_Mh_z0}
\end{figure*}

In our work, we take advantage of our knowledge of the HI content of model
galaxies both at the time the halo mass was maximum (we assume this
corresponds to the last time the galaxy was central), and at redshift 0. We
can thus verify the influence of the different satellite treatments on the
predicted relation between HI mass and halo mass.

In Fig.~\ref{fig:HI_Mh_ty0} we show the relation between the HI mass and the
halo mass at the last time the galaxy was central.  This figure represents what
we would find if we suppose the HI content of galaxies does not change after
accretion, as done in P13.  Solid lines show the median relation for all 
galaxies (black), centrals (blue) and today satellites (orange). The
1$\sigma$ spread is shown by the shaded area (for all galaxies) and the dashed
lines (for central and satellites). The relation obtained in P13 is
over-plotted in red (solid line is the median and dashed lines show the 
1$\sigma$ spread) for
a direct comparison. In this figure the relation obtained for satellite
galaxies is very close to that obtained for centrals, with some small
differences due to redshift evolution of the HI content of galaxies at the time
of accretion (see Fig.~\ref{fig:med_HIz_t}).

As already noted for the HI-$M_*$ relation (see Sec.~\ref{sec:HI_Ms}), B06
fails to reproduce the HI content in medium to high mass haloes because 
satellites are generally too gas poor and central galaxies are depleted of 
their cold gas by efficient radio mode feedback. The other 
models exhibit a wide scatter in the relation that is
almost independent of $M_{200}^{\rm max}$, with values 
of $\sigma_{\rm HI}\sim0.3-0.5$~dex depending on the
model. The shape is qualitatively similar to the one inferred in P13 if we
consider H15, GAEA and XBR16, while the other models predict more HI for 
$M_{200}^{\rm max} \lesssim 10^{12}\;M_{\sun}$, and less HI than that inferred 
by P13 in the most massive haloes. 

In Fig.~\ref{fig:HI_Mh_z0} we take into account the evolution of 
HI in satellites, that is we used for all galaxies their HI content at
redshift 0, while $M_{200}^{\rm max}$ corresponds again to the parent halo mass
at the last time the galaxy was central. 
In this figure we can appreciate a down-shift
of 0.1-0.4 dex of the relation for $M_{200}^{\rm max}<10^{12.5-13}\;
M_{\odot}$. The shift is driven by the satellites 
with lowest masses (both halo and HI), whose HI is depleted after
accretion because consumed through star formation.
The gas depletion increases the scatter of the total relation, with the effect
being enhanced for B06, DLB07 and GAEA.  

As noted in Sec.~\ref{sec:results_HI_satellites}, the lowest HI bin 
considered ($[10^{8.5};\;10^{9.5}]\;M_{\sun}$) is dominated by satellite galaxies, 
and therefore the one 
most affected by a different treatment for the evolution of these galaxies.

\subsection{HI galaxy content and halo spin}
\label{sec:results_HI_spin}

A relatively tight correlation is expected between the values of the halo spin
and the gas content of galaxies. \citet{huang2012} studied this relation
for the  
ALFALFA HI-selected
galaxies. The spin of the dark matter halo was calculated using the 
$\lambda$ estimator proposed by \citet{hernandez2007}, assuming
a dark matter isothermal density profile, an exponential surface density 
profile for the 
stellar disk, and a flat disk rotation curve. Using these assumptions:
$\lambda\propto r_{disk}/V_{rot}$, thus the spin depends on the 
disk scale radius and the rotational velocity. \citet{huang2012} found that 
ALFALFA HI rich galaxies favour high-$\lambda$ values.  \citet{kim_lee2013} 
simulated the evolution of dwarf-size haloes with varying halo-spin parameters 
and initial baryon
fractions, and found a correlation between disk radius (and therefore gas mass)
and $\lambda$. 

P13 used their measurements of the 2PCF to infer a relation between 
halo spin and gas content. They divided the haloes of the Bolshoi
simulation into three conveniently chosen spin bins: low spin with
$\lambda\in[0.002;0.02]$, medium spin with $\lambda\in[0.02;0.05]$ and high
spin, with $\lambda\in[0.05;0.20]$.  Assigning an HI content to each halo
according to the relation presented in Sec.~\ref{sec:HI_mhalo_max},
they found that the high and medium spin bins have almost the same 2PCFs 
measured for 
the HI selected galaxies, while low spin haloes have larger clustering signal. 
Based on these results, they argued that halo spin is the main driver of the HI
content of galaxies.
We have verified that selecting halos in the same range of spin values used 
by P13, we find results consistent with theirs.

Since the models we have used in our study are coupled with a high-resolution
cosmological simulation (MS), we can explicitly analyze the correlation between the
HI content of model galaxies and the spin of their parent haloes. For the B06 
model, information about
the spin is not available, so we exclude this model from the
following analysis. For central galaxies, we just use the spin of the parent
halo. For satellite galaxies, we consider the spin measured for the
parent halo at the last time they were central galaxies.

\begin{figure*}
\includegraphics[trim=1.5cm 2cm 2.5cm 1.0cm, clip, width = 0.6 \paperwidth]{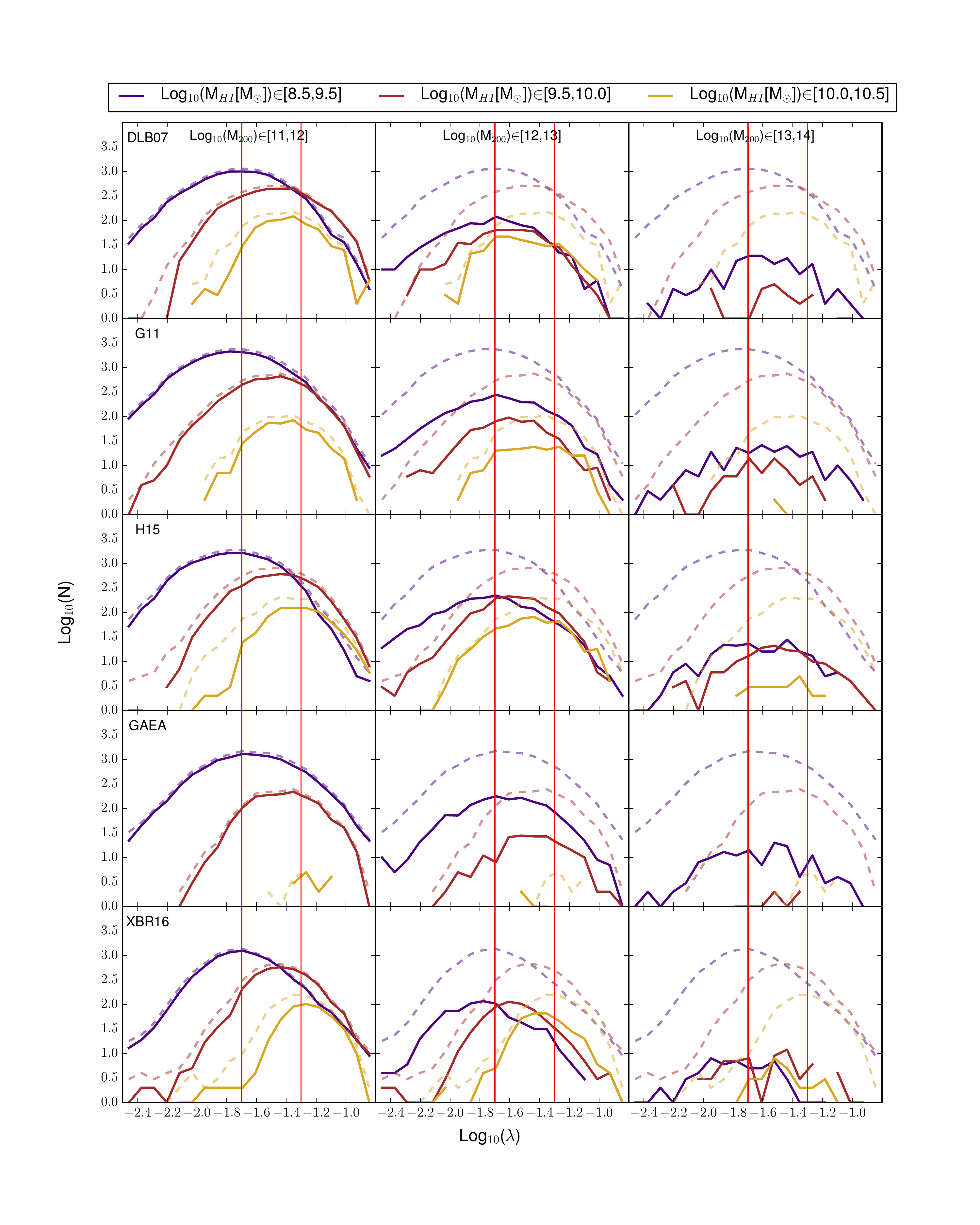}
  \caption{Distribution of halo spin values for galaxies selected in haloes
    with different 
    $M_{200}^{max}$ (columns) and HI mass (same colours of
    Fig.~\ref{fig:corr_HI_bin}), for all models used in our study (different
    rows).  The solid 
    lines represent the distribution for each selection, while dashed lines are
    the total distribution over all the $M_{200}^{max}$ range, reproduced in
    every panel as reference. The vertical red lines show the spin division
    considered by P13, and are plotted as reference.}
  \label{fig:spin_hist_HI_bin}
\end{figure*}

We show the histograms of the spin for some chosen $M_{200}^{\rm max}$ 
(different columns) and for the usual ranges in HI content in
Fig.~\ref{fig:spin_hist_HI_bin}.  
Dashed lines correspond to the distributions obtained for each HI mass, 
independently of $M_{200}^{\rm max}$, in order the easily compare each
$M_{200}^{max}$ bin to the total. The red vertical
lines correspond to the bins used in P13. 

The majority of the galaxies with the lowest HI mass ($[10^{8.5};\;10^{9.5}]\;M_{\sun}$)
are found in haloes 
with low spin values ($\lambda\in[0.002;0.02]$), 
while the majority of galaxies in the two HI richer bins 
($[10^{9.5};\;10^{10.5}]\;M_{\sun}$)
correspond to mid-spin values ($\lambda\in[0.02;0.05]$). 
Interestingly, this distribution is very 
similar for all models, i.e. the HI-spin correlation/distribution is not 
dependent on the specific prescriptions of each model. This result can be 
understood as follows: the halo spin parameter is used to set the initial 
value of the disk scale-length, and this quantity is then used to compute a 
density threshold for star formation. In the case of a low spin halo, the 
initial radius will be small, and the surface density  will be large. The star 
formation rate is directly dependent on the surface density and thus, for 
a fixed value of HI, a smaller radius will correspond to a faster HI depletion.
This explains why HI rich galaxies are rare in haloes with low spin values, and
why high spin (large disk radius) haloes host galaxies with a broad range of
HI masses. 

The figure shows that, in all models considered in our study, there is no
tight correlation between the HI content of galaxies and the spin of their 
parent haloes: low spin haloes are more likely populated by galaxies with 
low HI mass, and HI rich galaxies are most likely hosted by haloes with 
large spin values, but high spin haloes are populated by galaxies in a wide
range of HI mass.

\section{Conclusions}
\label{sec:conclusions}

In this work, we study the basic statistical properties of HI selected 
galaxies extracted from semi-analytic models of galaxy formation, and compare 
theoretical predictions with available data. In particular, we use four models 
whose galaxy catalogues are publicly available 
\citep{bower06,delucia07,guo10,henri15}, and two models recently developed by 
our group \citep[][and Xie et al. in preparation]{hirschmann2015}. Only one of 
these models (that described in Xie et al.) includes an explicit 
modelling for the partition of cold gas into atomic and molecular components, 
and a molecular hydrogen based star formation law.  All models are run on the 
same cosmological simulation, the Millennium Simulation \citep{springel2005}.

For those model in which the cold gas is treated as a single star forming phase,
we estimate the HI content of model galaxies {\it a posteriori} assuming 
\citep[as in][]{obreschkow2009} that the cold gas is distributed in an 
exponential disk and that the ratio between molecular and atomic hydrogen is 
determined by the kinetic gas pressure. All models used in our study include 
different specific modelling for the various physical processes considered and,
in particular, for the evolution of satellite galaxies. We find this to have 
relevant consequences on model predictions for HI selected galaxies. 

All models considered are in relatively good agreement with the observed local 
HI mass function, with the exception of the model by \citet{bower06} that predict 
too many galaxies with intermediate to large HI mass and too few galaxies with
small HI content ($M_{HI}<10^{9.6}M_{\sun}$). We find this is due to excessive HI masses in low stellar 
mass galaxies. This particular model also fails to reproduce the observed 
scaling relations as it predicts very little HI associated with satellites 
(likely because of instantaneous stripping of hot gas and efficient stellar 
feedback), and central galaxies 
with stellar masses larger than $\sim 3\times10^{10}\,{\rm M}_{\odot}$ (because
of too efficient radio mode feedback). The  
observed scaling relations are relatively well reproduced by all other models,
but some of them \citep[][and Xie et al.]{guo10, hirschmann2015} exhibit a deficit 
of HI rich galaxies ($M_{HI}>10^{10}M_{\sun}$) at any stellar mass. 

The HI content of satellite galaxies varies significantly among the models 
considered, due to a different treatment for the hot gas associated with 
infalling galaxies and stellar feedback. As expected, models that assume a non 
instantaneous stripping of this hot gas reservoir tend to predict larger HI 
masses for satellite galaxies. The most massive satellites, in particular, 
tend to have an average HI content that is very close to that of central 
galaxies of the same stellar mass. This is due to the fact that these galaxies 
were accreted relatively recently and evolved as central galaxies for most of 
their lifetime \citep{delucia12}. Interestingly, assuming an instantaneous 
stripping of the hot gas at the time of accretion, as in GAEA, does not 
necessarily imply low HI content for 
satellite galaxies. As already noted in \citet{hirschmann2015}, this model is 
characterized by significantly lower fractions of passive (and 
therefore larger fractions of gas-rich, star forming) satellites with respect to
e.g. the model presented in \citet{delucia07}. This is a consequence of 
suppressed and delayed star formation at early times, and leads to larger cold gas
fractions at the time of accretion. 

Using galaxy catalogues from each model, we have built mock light-cones that
we have used to analyze how the clustering of HI selected galaxies compares to 
recent measurements by \citet{papastergis2013}. In particular, we have 
considered three HI mass bins: low 
($M_{\rm HI}\in[10^{8.5};\;10^{9.5}]\;M_{\sun}$), intermediate 
($M_{\rm HI}\in[10^{9.5};\;10^{10.}]\;M_{\sun}$) and high 
($M_{\rm HI}\in[10^{10};\;10^{10.5}]\;M_{\sun}$). The lowest HI mass bin is likely 
affected by limited sampling volume in the observations so, although the data
suggest a lower clustering signal for this particular bin, \citet{papastergis2013}
argue that this is not significant and that the 2-point clustering function 
measured for these three bins are not statistically different. In contrast, we
find that all models predict for galaxies in the lowest HI bin a clustering 
signal {\it higher} than for the HI richer galaxies. For the other two bins 
($[10^{9.5};\;10^{10.5}]\;M_{\sun}$),
half of the models are in relatively good agreement with data while the other
half tend to under-predict slightly the measured clustering. Interestingly, 
the model by \citet{bower06} that has the worst performance for the HI mass
distribution and scaling relations, exhibits the best 
agreement with data by \citet{papastergis2013} in the lowest HI bin, with a clustering signal  only 
slightly stronger than that for HI richer galaxies. We show that the lowest
HI bin is strongly affected by the adopted treatment for satellite galaxies as
this bin is dominated by this galaxy population.
 
The relation between the HI mass (at the accretion time for satellite galaxies)
and halo mass (at its maximum) predicted by all models considered is in quite
good agreement with that inferred by \citet{papastergis2013}. Again, the
exception is the model by \citet{bower06} that predicts negligible HI in haloes
more massive than $\sim 10^{12}\,{\rm M}_{\odot}$. This is interesting because,
as noted above, the clustering signal predicted by this particular model is the
one that is closest to observational measurements. Thus, taken at face value,
these results suggest that the clustering of HI selected galaxies does not
provide enough information to constrain the relation between halo mass and HI
mass, and that a crucial element is represented by the evolution of the HI
content of satellite galaxies. The scatter of the predicted relation increases
in case one considers the HI associated with galaxies at present time, because
of gas depletion in satellite galaxies. Specifically, we find that the
\citet{bower06} model exhibits the shortest gas consumption times: a galaxy
accreted at $z \sim 1$ with HI mass $\sim 10^{10}\,{\rm M}_{\odot}$ conserves
only about $10$ per cent of this gas after $2$~Gyr ($1$ per cent after $3$
Gyr). This is likely due to an efficient stellar feedback, coupled with an
instantaneous stripping of the hot gas associated with infalling galaxies.  The
gas consumption timescale is typically longer in the other models, and the
assumption of a gas density threshold for star formation implies that the gas
associated with satellite galaxies never falls below such limit. The gas
consumption timescales are longest in the Xie et al. model where star formation
is evaluated in radial annuli, considering the local physical conditions of the
inter-stellar medium.

Finally, we have examined the relation between the HI content of galaxies and 
the spin of the parent dark matter halo (we have considered the value at the 
time of accretion for satellite galaxies). We find that low spin haloes 
($\lambda\in[0.002;0.02]$) are more  
likely populated by HI poor galaxies ($[10^{8.5};\;10^{9.5}]\;M_{\sun}$), 
and HI richer galaxies tend to reside
in haloes with large spin ($\lambda\in[0.02;0.2]$). The scatter, however, 
is relatively large and 
haloes with intermediate and large spin values tend to host galaxies with a 
large dynamic range in HI mass. Interestingly, the distributions are very 
similar in all models considered, i.e. they are not significantly affected by
the specific modelling of 
the various physical processes affecting the HI 
content of galaxies. 
This is somewhat 
surprising as the spin enters the calculation of the disk radius and this, in
turn, affects the star formation rate (and therefore the gas content). In most
models, however, the halo spin is only used to determine the initial radius
of the gaseous component and does not affect significantly the subsequent 
evolution. The dependence of the gas initial disk radius on halo spin explains
the trends found: an initial 
small radius and a large gas fraction translates into a high surface density
and therefore into a large star formation rate, which consumes the gas
rapidly. This 
explains why haloes with small spin values tend to be associated with gas
poor galaxies. 

Our analysis show that different models lead to very similar results for
galaxies with intermediate to large HI mass ($[10^{9.5};\;10^{10.5}]\;M_{\sun}$), 
while significant differences can be found for relatively HI poor 
galaxies ($[10^{8.5};\;10^{9.5}]\;M_{\sun}$). As discussed above, this bin is
dominated by satellite galaxies and therefore mostly affected by the different
treatment for this particular galaxy population. More detailed data in
this particular HI mass range are needed to put stronger constraints on galaxy
formation models.  Dedicated controlled simulations would be
useful to quantify the effect of stripping processes in satellites \citep[see
e.g.][]{Tonnesen_etal_2009}.  
Some attempts in this direction were made using
semi-analytic models based on Monte Carlo merger trees 
\citep{lagos2011mol,kim_hs2015} 
or on the Millennium Simulation II \citep{kim_hs2016}, 
and have shown that these models can be
used to constrain the physics of low mass satellites.

At the same time, we need larger statistical
samples and
the possibility to estimate the galaxy `hierarchy' (i.e. being a satellite or a
central) in observations.  New radio instruments, such as
SKA\footnote{https://www.skatelescope.org/project/} and its precursors
\citep[][for ASKAP and MeerKAT, respectively]{johnston2008,booth2009}, will 
provide valuable data for these analyses.

\section*{Acknowledgements}
The Millennium Simulation databases used in this paper and the web application
providing online access to them were constructed as part of the activities of
the German Astrophysical Virtual Observatory (GAVO). This work has been
supported by the MERAC foundation and by the PRIN-INAF 2012 grant `The Universe
in a Box: Multi-scale Simulations of Cosmic Structures'. We acknowledge a
CINECA award under the ISCRA initiative, for the availability of high
performance computing resources and support. Michaela Hirschmann acknowledges
financial support from the European Research Council via an Advanced Grant
under grant agreement no. 321323 NEOGAL.




\bibliographystyle{mnras}
\bibliography{biblio} 

%
%
%
%
%

\bsp	
\label{lastpage}
\end{document}